\documentclass[showpacs,preprintnumbers,amsmath,amssymb,APSl,prd,nofootinbib,superscriptaddress]{revtex4-2}
\usepackage{graphicx,color}
\usepackage{amsmath}
\usepackage{amssymb}
\usepackage{hyperref}
\usepackage[utf8]{inputenc}
\usepackage[english]{babel}
\usepackage{epsfig}
\usepackage{subfigure}
\usepackage{wasysym}
\usepackage{color,xcolor}
\usepackage{amsmath}
\usepackage{bm}
\usepackage{epsfig}
\usepackage{amsfonts}
\usepackage{dcolumn}
\usepackage{float}

\hypersetup{colorlinks=true, linkcolor=blue, citecolor=green}

\usepackage{dcolumn}
\usepackage{bm}
\usepackage{ifpdf}
\usepackage{hyperref}
\usepackage{float}
\usepackage{bm}
\usepackage{xcolor,color,graphicx,graphics}
\usepackage[OT1]{fontenc}
\usepackage{latexsym,amssymb,amsmath,amsfonts}
\usepackage{makeidx}
\usepackage{epsfig}
\usepackage{epstopdf}
\usepackage{mathrsfs}
\hypersetup{colorlinks=true, linkcolor=blue, citecolor=green}
\usepackage{enumerate}
\usepackage{xcolor}
\usepackage{multirow}

\def\nn{\nonumber\\}
\newcommand{\f}[2]{\frac{#1}{#2}}
\def\be{\begin{equation}}
	\def\ee{\end{equation}}
\def\bea{\begin{eqnarray}}
	\def\eea{\end{eqnarray}}

\def\bwt{\begin{widetext}}
	\def\ewt{\end{widetext}}

\usepackage{amsmath}

\usepackage{amssymb}

\begin{document}
\title{Trace anomaly contributions to semi-classical wormhole geometries}

\author{Mohammad Reza Mehdizadeh} 
\email{mehdizadeh.mr@uk.ac.ir}        
\affiliation{Department~of~ Physics,~ Shahid~ Bahonar~ University, P.~ O.~ Box~ 76175, Kerman, Iran}

\author{Amir Hadi Ziaie} \email{ah.ziaie@riaam.ac.ir}
\affiliation{Research~Institute~for~Astronomy~and~Astrophysics~of~ Maragha~(RIAAM), University of Maragheh,  P.~O.~Box~55136-553,~Maragheh, Iran}

\author{Francisco S. N. Lobo} \email{fslobo@ciencias.ulisboa.pt}
\affiliation{Instituto de Astrof\'{i}sica e Ci\^{e}ncias do Espa\c{c}o, Faculdade de Ci\^{e}ncias da Universidade de Lisboa, Edifício C8, Campo Grande, P-1749-016 Lisbon, Portugal}
\affiliation{Departamento de F\'{i}sica, Faculdade de Ci\^{e}ncias da Universidade de Lisboa, Edif\'{i}cio C8, Campo Grande, P-1749-016 Lisbon, Portugal}


\begin{abstract}
We investigate wormhole solutions within the framework of the semi-classical Einstein equations in the presence of the conformal anomaly (or trace anomaly). These solutions are sourced by a stress-energy tensor (SET) derived from the trace anomaly, and depend on two positive coefficients, $\alpha$ and $\lambda$, determined by the matter content of the theory and on the degrees of freedom of the involved quantum fields. For a Type B anomaly ($\alpha=0$), we obtain wormhole geometries assuming a constant redshift function and show that the SET components increase with the parameter $\lambda$. In the case of a Type A anomaly ($\lambda=0$), we generalize previously known solutions, yielding a family of geometries that includes Lorentzian wormholes, naked singularities, and the Schwarzschild black hole. Using isotropic coordinates, we identify parameter choices that produce traversable wormhole solutions. Extending to the full trace-anomaly contribution, we solve the differential equation near the throat to obtain the redshift function and demonstrate that both the Ricci and Kretschmann scalars remain finite at the throat. We further analyze the trajectories of null and timelike particles, showing that the height and width of the effective potential for null geodesics increase monotonically with $\alpha$, while the innermost stable circular orbit (ISCO) radius also grows with larger $\alpha$. These results illustrate the rich interplay between trace anomaly effects and the structure and dynamics of wormhole spacetimes.
\end{abstract}

\date{\today}

\maketitle


\section{Introduction}

Wormholes are hypothetical structures that act as bridges connecting either distinct universes or distant regions within the same universe. The concept was first introduced in 1935 by Einstein and Rosen in the form of the so-called Einstein–Rosen bridge~\cite{ERB}. The term \textit{wormhole} itself was later coined by Misner and Wheeler in 1957, in the context of their study on geon-like configurations within general relativity (GR)~\cite{mwhee}. Their analysis demonstrated that wormholes linking asymptotically flat spacetimes could serve as non-trivial solutions to the Einstein–Maxwell equations, allowing electric field lines to pass through the wormhole throat and emerge in another region of spacetime. 
A major development occurred in 1988 when Morris and Thorne examined \emph{traversable wormholes}—spacetime geometries permitting two-way travel through a minimal surface, known as the throat~\cite{mothorn}. Their results showed that such configurations inevitably require \emph{exotic matter} that violates the null energy condition (NEC)~\cite{book visser,kar1}. The necessity of exotic matter remains a central challenge in wormhole physics, motivating research into strategies for minimizing or avoiding such violations~\cite{pow1}. 

Several proposals address this issue. Dynamical wormholes, for instance, can satisfy the energy conditions temporarily along certain geodesics~\cite{dynami1}. Thin-shell wormholes, constructed via the cut-and-paste technique, confine exotic matter to a thin layer at the throat~\cite{pvis}. Other approaches invoke phantom energy, quintom fields, or interactions between dark matter and dark energy as supporting sources~\cite{phantworm,intdarksec}. For comprehensive overviews, we refer the reader to Refs.~\cite{hisnotewo,loboreview}.
Modified theories of gravity provide yet another route, wherein higher-order curvature terms can allow the existence of wormholes supported by ordinary matter \cite{Lobo:2009ip,Harko:2013yb}. Notable results have been obtained in frameworks such as Einstein–Gauss–Bonnet gravity, Lovelock gravity, and $f(R)$ gravity, among others~\cite{highthin,EGBW,highdgr,nonsymgr,loveworm,frworm,frtworm,bdw,othmodw}. These theories frequently minimize or remove the need for NEC violation, thereby opening new possibilities for the theoretical construction of traversable wormholes~\cite{mothorn,highthin,frworm}. 

Beyond classical and modified gravity frameworks, quantum corrections can also play a decisive role in wormhole physics. In particular, effects such as the conformal anomaly, or \emph{trace anomaly}, can yield novel wormhole solutions in which semi-classical phenomena replace the role of exotic matter, thereby offering a natural bridge between classical and quantum regimes~\cite{fabr1,fabr2,fabr2a,fabr2b}. The trace anomaly arises when a classically conformally invariant field theory fails to preserve conformal symmetry at the one-loop level upon quantization, resulting in a non-zero trace of the renormalized SET, $\langle T^{\mu}{}_{\mu} \rangle$~\cite{des1997,des1997a,des1997b,des1997c,des1997d,bet1a}.
Indeed, the trace anomaly plays a fundamental role in semi-classical gravity and has been extensively investigated over several decades~\cite{transubject,transubject1,transubject2,transubject3,transubject4,transubject5}. As demonstrated in~\cite{cai1}, the anomaly modifies Einstein’s field equations and significantly affects the spacetime geometry in the vicinity of singularities and high-curvature regions. 

In cosmological contexts, the trace anomaly can act as a driving mechanism for inflationary expansion or as an effective source of dark energy. Notably, Starobinsky’s inflationary model can be derived from anomaly-induced quantum corrections~\cite{Hawkinfl,bamba2001}. Likewise, in late-time cosmology, contributions from the anomaly may mimic dark energy, thereby influencing the Universe’s expansion history~\cite{darkanom0,darkanom,darkanoma}. 
In black hole physics, the trace anomaly has profound effects on the properties of event horizons and the strucutre of singularities \cite{fabr1,fabr2,fabr2a,fabr2b,casa,Ho2018,Cai2014,Anderson2007}. For example, semi-classical corrections to Einstein’s equations can deform the horizon of a classical Schwarzschild black hole into a wormhole-like configuration without an event horizon~\cite{casa,casaa,casab,casac,casad,casaf}. Comparable phenomena occur in the case of rotating black holes, where the anomaly influences the ergoregion dynamics and contributes to Hawking radiation \cite{Fernandes2023,Balbinot1999,TekinAbedi,TekinAbedi1}.

Furthermore, in the study of semi-classical relativistic stars, the condition of stellar equilibrium is governed by a generalized extension of the classical Tolman–Oppenheimer–Volkoff (TOV) equation~\cite{TOV,TOV1}. This generalized form emerges naturally once quantum corrections from semi-classical gravity are taken into account. Among the various sources of such corrections, the trace anomaly is particularly important, as it introduces non-trivial modifications to the equilibrium conditions through additional contributions to both the effective pressure and the energy density of the stellar matter. 
These corrections lead to a refined framework that extends the classical description of compact stellar objects, thereby capturing the interplay between relativistic gravitational effects and quantum field-theoretic phenomena. As a result, the semi-classical TOV equation provides a more accurate and self-consistent characterization of equilibrium configurations, especially in regimes where extreme densities and curvatures render purely classical treatments insufficient. In this sense, the semi-classical approach effectively bridges the conceptual and mathematical gap between general relativity and quantum field theory in curved spacetime~\cite{SSTRAN,SSTRAN1,SSTRAN2,SSTRAN3,SSTRAN4}.

In the context of wormhole physics, anomaly-induced modifications have been recognized as a plausible source of the negative-pressure fluid required to violate the NEC~\cite{casa}. In this context, the possibility of generating primordial, spherically symmetric wormholes during the early stages of the Universe, under the influence of anomaly effects, has been analyzed in detail in~\cite{anomwormpreun,anomwormpreun1}. These studies highlight the potential role of quantum corrections in shaping the formation and stability of wormhole-like structures in a cosmological setting.  
Recent progress in holographic gravity, particularly within the AdS/CFT correspondence framework, further suggests that the boundary trace anomaly may impose non-trivial constraints on bulk wormhole geometries. This insight establishes an explicit connection between quantum field theory effects at the boundary and the question of wormhole traversability in the bulk spacetime~\cite{wormadscft,wormadscft1}.  

Moreover, advances in non-perturbative approaches to quantum gravity, most notably the asymptotic safety program, indicate that anomaly-induced contributions could play a stabilizing role for Lorentzian wormholes. Remarkably, this stabilization mechanism may operate without invoking exotic matter sources, which are usually deemed necessary within the classical framework~\cite{wormassafe,wormassafe1,wormassafe2,wormassafe3,wormassafe4,wormassafe5}. Indeed, significant advances in non-perturbative approaches to quantum gravity, with the asymptotic safety program serving as a prominent example, suggest that anomaly-induced contributions may play a decisive role in stabilizing Lorentzian wormholes. In contrast to the classical picture, where exotic matter is typically required to maintain traversability, these quantum corrections open the possibility of stable wormhole configurations without the need to invoke such exotic sources~\cite{wormassafe,wormassafe1,wormassafe2,wormassafe3,wormassafe4,wormassafe5}. 
These results represents a substantial conceptual shift, as it indicates that the quantum structure of spacetime itself could naturally provide the conditions necessary for wormhole stability. Taken together, these findings highlight the central importance of the trace anomaly, as a unifying element that bridges semi-classical, holographic, and non-perturbative perspectives in wormhole physics.

Guided by these motivations, this work focuses on exploring a novel class of traversable wormhole solutions within the framework of the semi-classical Einstein equations incorporating trace-anomaly effects. The corresponding spacetimes are supported by a SET arising from the trace anomaly, characterized by two positive parameters, $\alpha$ and $\lambda$, which encode the matter content of the theory and the degrees of freedom of the underlying quantum fields. By examining both Type A and Type B anomalies, we construct a variety of geometries, encompassing Lorentzian wormholes, naked singularities, and Schwarzschild black holes, and determine the ranges of $\alpha$ and $\lambda$ that produce physically consistent solutions. We find that, at the wormhole throat, curvature invariants such as the Ricci and Kretschmann scalars remain finite, while the components of the SET and associated pressures are sensitive to the anomaly parameters. Additionally, the study of particle trajectories reveals that the effective potential and innermost stable circular orbit (ISCO) radius grow with increasing $\alpha$, demonstrating that trace-anomaly contributions play a significant role in shaping both the geometry of the wormhole and the dynamics of test particles within it.

The paper is organized as follows: In Sec.~\ref{Gb}, we provide a concise overview of wormhole geometries supported by trace-anomaly contributions, highlighting the role of the anomaly parameters $\alpha$ and $\lambda$ and their influence on the SET. In Sec.~\ref{WHS}, we construct new wormhole solutions by considering both constant and radially varying redshift functions, and we discuss the resulting spacetime structures, including Lorentzian wormholes, naked singularities, and Schwarzschild black holes. In Sec.~\ref{WHtrajectory}, we investigate the motion of null and timelike particles in these wormhole spacetimes, examining the effective potential and the innermost stable circular orbits (ISCOs), and analyzing how these quantities are affected by the trace-anomaly parameters. Throughout this work, we adopt natural units by setting $\hbar = c = G = 1$.

\section{Trace anomaly wormhole geometries}\label{Gb}

\subsection{Trace anomaly}

The trace anomaly is a quantum effect arising from the contribution of (massless) matter fields. While the classical field action in curved spacetime is conformally invariant, renormalization is required to eliminate divergences originating from one-loop vacuum contributions. Consequently, the counter-terms introduced to remove these divergent poles break the conformal invariance of the matter action. Classically, the trace of the SET vanishes in a conformally invariant theory; however, after renormalization, a non-zero trace emerges, resulting in an anomalous SET. This phenomenon is known as the quantum anomaly or trace anomaly~\cite{des1997,des1997a,des1997b,des1997c,des1997d}. If we consider the back-reaction of the quantum fields to a curved spacetime geometry in the field equations of GR we get the semi-classical Einstein equation with a modified source as~\cite{darkanom0,bir,park,Hu2020}
\begin{equation}
G_{\mu\nu}= 8\pi\left(\langle T_{\mu\nu} \rangle + T^{\rm  class}_{\mu\nu}\right),
	\label{semiclassEFE}
\end{equation}
where $ T^{\rm class}_{\mu\nu}$ represents some classical gravitational source, while $\langle T_{\mu\nu} \rangle$ denotes the expectation value of the SET, an effective SET which originates from quantum loops and must be covariantly conserved, i.e., it obeys the continuity equation $\nabla^{\mu} \langle T_{\mu\nu} \rangle=0$. 

It should be noted that, deriving the explicit form of this tensor for a generic spacetime is almost impossible, necessitating simplifying assumptions about the background geometry~\cite{bir,park,Hu2020}. However, in four dimensions, the trace anomaly takes the well-known form
 \begin{align}
\langle T\rangle={\tilde{\lambda}}{\cal C}^{2} - {\tilde{\alpha}}{\cal G}, \label{excep1}
 \end{align}
where, $T=g^{\mu\nu}T_{\mu\nu}$ is the trace of the SET and the scalars ${\cal G}$ and ${\cal C}^2$ are the Gauss-Bonnet four-dimensional topological invariant and the square of Weyl tensor, respectively. These scalars are given by  
\begin{equation}
{\cal G}=R^2-4 R_{\mu\nu}R^{\mu\nu} +  R_{\mu\nu\alpha\beta}R^{\mu\nu\alpha\beta}\,,
\qquad
{\cal C}^2\equiv C_{\mu\nu\alpha\beta}C^{\mu\nu\alpha\beta}=\frac{R^2}{3}-2 R_{\mu\nu}R^{\mu\nu} +  R_{\mu\nu\alpha\beta}R^{\mu\nu\alpha\beta}\,,
  \end{equation}
respectively, where $R_{\mu\nu\alpha\beta}$ is the Riemann tensor. The first term in the right-hand-side of Eq.~(\ref{excep1}) is called a type B anomaly, while the second one is called type A anomaly~\cite{deser}. The coefficients ${\tilde{\alpha}}$ and ${\tilde{\lambda}}$ are two positive constants determined by the matter content of the theory and depend on the degrees of freedom of the involved quantum fields. These coefficients are given by
\begin{align}
{\tilde{\lambda}}=\frac{1}{120(4\pi)^2}\left[n_0+6n_{1/2}+12n_1\right], \qquad {\tilde{\alpha}}=\frac{1}{360(4\pi)^2}\left[n_0+11n_{1/2}+62n_1\right],
\end{align}
respectively, where $n_0$, $n_{1/2}$ and $n_1$ are the number of massless fields of spin $\left(0,1/2,1\right)$, respectively, in the conformal field theory. 

\subsection{Morris-Thorne wormhole geometries: Semi-classical approach}

Now, in the context of wormhole physics, in their seminal work \cite{mothorn}, Morris and Thorne introduced the following spherically symmetric line element
\begin{equation}
ds^{2}=-e^{2\Phi (r)}dt^{2}+\frac{dr^{2}}{1-b(r)/r}+r^{2}\left(d\theta^{2} + \sin^2\theta \, d \phi^2 \right)\,, \label{mor1}
\end{equation}
as a possible solution to obtain viable traversable wormhole geometries. 
In the metric presented above, $\Phi(r)$ denotes the \emph{redshift function}, which determines the gravitational redshift experienced by signals traversing the spacetime. The function $b(r)$ is referred to as the \emph{wormhole shape function}, as it characterizes the spatial geometry of the wormhole and determines its spatial embedding.  The radial coordinate ranges from $r_0$ to spatial infinity where the surface at $r = r_0$ is known as the wormhole’s throat. Every wormhole configuration must have a minimum radius, i.e. at the throat of the wormhole $b(r_0) = r_0$, where $r_0$ being the minimum value of $r$. For the wormhole to possess a physically viable geometry, the shape function must satisfy the \emph{flaring-out condition} at the throat. Specifically, at the throat radius $r_0$, one must have $b'(r_0) < 1$, and throughout the entire spacetime region $r > r_0$, the condition $b(r) < r$ must hold. These requirements ensure that the wormhole throat is properly sustained and that the spacetime remains free from horizons or singularities in the traversable region.

 Note that the wormhole geometry generally requires two coordinate patches, each covering the domain \( r \in [r_0,+\infty[ \). Each patch corresponds to one of the two universes, and the patches meet smoothly at \( r=r_0 \), the location of the wormhole throat~\cite{book visser}. While the metric component \( g_{rr} = (1 - b(r)/r)^{-1} \) diverges as \( r \to r_0 \), this divergence is purely a coordinate effect. One may instead introduce the proper radial distance
\begin{equation}
l(r)=\pm\int_{r_0}^{r}\frac{dr}{\sqrt{1-b(r)/r}},
\label{prrad}
\end{equation}
which remains finite for all finite \( r \) throughout the spacetime, so that no thin-shell construction is required at \( r_0 \) (although it may be present~\cite{book visser}). The two signs correspond to the two asymptotically flat regions of the manifold: as \( l \to \pm \infty \) (equivalently \( r \to \infty \)), the ratio \( b(r)/r \to 0 \), demonstrating that the geometry approaches asymptotic flatness on both sides of the throat.

Considering now an orthonormal reference frame defined as
\bea\label{orthoreff}
{\rm e}_{\hat{t}}={\rm e}^{-\Phi}{\rm e}_{t}, \qquad {\rm e}_{\hat{r}}=\left(1-\f{b}{r}\right)^{\f{1}{2}}{\rm e}_{r}, \qquad {\rm e}_{\hat{\theta}}=\f{{\rm e}_{\theta}}{r}, \qquad {\rm e}_{\hat{\phi}}=\f{{\rm e}_{\phi}}{r\sin\theta},
\eea
the non-vanishing components of Einstein tensor in this frame are found as
\bea\label{gortho}
&&G_{\hat{t}\hat{t}}=\f{b^\prime}{r^2}, \qquad G_{\hat{r}\hat{r}}=-\f{b}{r^3}+2\left(1-\f{b}{r}\right)\f{\Phi^\prime}{r},
	\nn
&&G_{\hat{\theta}\hat{\theta}}=G_{\hat{\phi}\hat{\phi}}=\left(1-\f{b}{r}\right)\left[\Phi^{\prime\prime}+(\Phi^{\prime})^2-\f{rb^\prime-b}{2r(r-b)}\Phi^\prime-\f{rb^\prime-b}{2r^2(r-b)}+\f{\Phi^\prime}{r}\right],
\eea
where a prime denotes differentiation with respect to $r$. 

Throughout this work we consider $T^{\rm class}_{\mu\nu}=0$. Thus, the nonzero components of the SET in the orthonormal reference frame are given by
\begin{equation}\label{setortho}
\langle T_{\hat{t}\hat{t}}\rangle=\rho(r), \qquad \langle T_{\hat{r}\hat{r}}\rangle=p_r(r), \qquad \langle T_{\hat{\theta}\hat{\theta}}\rangle=\langle T_{\hat{\phi}\hat{\phi}} \rangle=p_t(r),
\end{equation}
where $\rho(r)$ is the energy density and $p_{r}(r)$ and $p_{t}(r)$ are the radial and transverse pressures, respectively. 

Furthermore, for notational simplicity, we define the positive metric functions $f(r)$ and $g(r)$ as
\bea
f(r)\equiv e^{2\Phi(r)}, \qquad g(r)\equiv  1-\frac{b(r)}{r},
\label{fgfuncs}
\eea  
respectively, so that the semi-classical Einstein field equation (\ref{semiclassEFE}) provides us with the following expressions
\bea
8\pi r^2\rho(r)&=&1-g^{\prime}(r)r-g(r),\label{feild1}\\
8\pi r^2p_{r}(r)&=&rg(r)\frac{f^{\prime}(r)}{f(r)}+g(r)-1,\label{feild2}\\
32\pi r^2p_{t}(r)&=&2rg^\prime(r)-g(r) \left[\f{rf^\prime(r)}{f(r)}\right]^2+r^2g^\prime(r)\f{f^\prime(r)}{f(r)}+\f{2rg(r)}{f(r)}\,\left[f^\prime(r)+rf^{\prime\prime}(r)\right].
	\label{feild3}
\eea

\subsection{Energy conditions}

Using the metric functions (\ref{fgfuncs}), the flaring–out condition at the wormhole throat requires that $g(r_0) = 0$ and $g'(r_0) > 0$, where $r_0$ denotes the throat radius.  
It is well known that the existence of traversable Lorentzian wormholes in four dimensions, as solutions of the Einstein field equations, necessarily involves some form of so-called ``exotic matter'', i.e., matter violating the NEC~\cite{kar1}, as mentioned above.  
This violation arises from the requirement that the flaring-out condition be satisfied both at $r = r_0$ and in its immediate neighborhood.
Physically, the flaring-out condition ensures that the wormhole throat does not collapse, making it essential for traversability.  
In classical GR, such a configuration inevitably requires exotic matter localized at, or near, the throat.
The NEC is a special case of the weak energy condition (WEC), which asserts that the energy density measured by any observer is non-negative.  
More specifically, the WEC demands that $T_{\mu\nu} U^\mu U^\nu \geq 0$ for any timelike vector field $U^\mu$.  
For the SET given in Eq.~(\ref{setortho}), the WEC reduces to the following inequalities:
\begin{equation}\label{E11}
\rho \geq0 , \qquad \rho+p_{r}\geq0, \qquad \rho+p_{t}\geq0.
\end{equation} 
Note that the last two inequalities are defined as the NEC~\cite{book visser}. Using Eqs.~(\ref{feild1})-(\ref{feild3}), one finds the following relationships
\bea
&&8\pi(\rho +p_{r})=\f{1}{r}\left[g(r)\f{f^{\prime}(r)}{f(r)}-g^{\prime}(r)\right],\label{EGBNEC}\\
&&32\pi(\rho +p_{t})=2g(r)\f{f^{\prime\prime}(r)}{f(r)}+\f{f^\prime(r)}{rf(r)}\left[2g(r)+rg^\prime(r)\right]-g(r)\left[\f{f^\prime(r)}{f(r)}\right]^2\!\!-\f{2}{r^2}\left[rg^\prime(r)+2g(r)-2\right]\,,
\label{EGBNEC1}
\label{EGBNEC2}
\eea
which reduce to the following expressions at the throat:
\begin{eqnarray}
\rho + p_r\Big|_{r=r_0}&=&-\f{g^\prime(r_0)}{8\pi r_0},\nn32\pi(\rho + p_t)\Big|_{r=r_0}&=&\f{g^\prime(r_0)f^\prime(r_0)}{f(r_0)}-\f{2}{r_0}g^\prime(r_0)+\f{4}{r_0^2}\,.
\end{eqnarray}
From this, we see that the NEC in the radial direction is violated as a direct consequence of the flaring–out condition.  
In contrast, in the tangential direction, whether the NEC is satisfied depends on the specific values of the metric components and their derivatives at the throat.

\subsection{Ricci and Kretschmann scalars: Regularity of the solutions}

Furthermore, for the line element~(\ref{mor1}), with the metric functions (\ref{fgfuncs}), the Ricci scalar, defined as $\mathcal{R} = g_{\mu\nu}\mathcal{R}^{\mu\nu}$, and the Kretschmann scalar, defined as $\mathcal{K} = \mathcal{R}_{\mu\nu\alpha\beta}\mathcal{R}^{\mu\nu\alpha\beta}$, are given by
\begin{eqnarray}
\mathcal{R}(r)=g(r)\left[\f{f^\prime(r)}{f(r)}\right]^2-\left[4g(r)+rg^\prime(r)\right]\f{f^\prime(r)}{rf(r)}-2g(r)\f{f^{\prime\prime}(r)}{f(r)}-\f{4}{r^2}\left[g(r)+rg^\prime(r)-1\right],
\label{ric1}
\end{eqnarray}
\begin{eqnarray}
	\mathcal{K}(r) &=& \left[\f{g(r)f^{\prime\prime}(r)}{f(r)}\right]^2-\f{f^{\prime\prime}(r)}{f(r)}\left[\f{g(r)f^{\prime}(r)}{f(r)}\right]^2+g(r)g^{\prime}(r)\f{f^{\prime}(r)f^{\prime\prime}(r)}{f^2(r)}+\f{g(r)^2}{4}\left[\f{f^{\prime}(r)}{f(r)}\right]^4
	\nn
	&&-\f{1}{2}g(r)g^\prime(r)\left[\f{f^\prime(r)}{f(r)}\right]^3+\left[\f{f^\prime(r)g^\prime(r)}{2f(r)}\right]^2+2\left[\f{g(r)f^\prime(r)}{rf(r)}\right]^2+\f{4}{r^4}\left[\f{1}{2}\left(rg^\prime(r)\right)^2+g(r)^2-2g(r)+1\right],
	\label{kr11}
\end{eqnarray}
respectively.
%
The above two expressions are useful in determining the possible occurrence (or absence) of spacetime singularities through their divergent (regular) behaviors~\cite{RicKretch}.

\section{Specific wormhole solutions}\label{WHS}

\subsection{Specific case: constant redshift function}\label{WHS1}

In order to theoretically construct a wormhole, one may specify the shape function and/or the redshift function. In the present analysis, we take the trace anomaly, $\langle T^{\mu}{}_{\mu} \rangle$, to be given by
\begin{align}
\langle T^{\mu}{}_{\mu} \rangle=-\rho(r)+p_r(r)+2p_t(r).\label{costr1}
\end{align}
The above constraint translates into a condition involving the shape and redshift functions, along with their derivatives.
More specifically, we substitute the components of the SET from Eqs. (\ref{feild1})--(\ref{feild3}) into the right-hand-side of Eq. (\ref{costr1}). Then, by using Eq. (\ref{excep1}) for the left-hand-side of the Eq. (\ref{costr1}), we obtain the following relation:
\begin{eqnarray}
&\left[g^{2}\left(48\,{r}^{2}{f}^{3}{f''}-24{r}^{2}{f}^{2}{{f'}}^{2}\right)-24\left(2{r}^{2}{f}^{3}{f''}-3{f}^{3}{f'}{g'}{r}^{2}-{r}^{2}{f}^{2}{{f'}}^{2}\right) g\right]\alpha\notag \\&-24\alpha {f}^{3}{f'}{g'}{r}^{2}+\lambda\left(2{{f'}}^{2}{r}^{2}+4 rf{f'}-4{r}^{2}f{f''}-8{f}^{2}\right)^2g^{2}+\lambda\left( 4{f}^{2}\left(r{g'}+2 \right)-2{r}^{2}f{f'}{g'}\right)^{2} \notag \\&-\lambda \left(-2{f}^{2}\left(r{g'}+2\right)+{r}^{2}f{f'}{g'}\right)\left( -2{{f'}}^{2}{r}^{2}-4rf{f'}+4{r}^{2}f{f''}+8{f}^{2} \right) g+24r^3f^4 g^{\prime} \notag \\&+{r}^{2}{f}^{2} \left(-6{{f'}}^{2}{r}^{2}+24rf{f'}+{r}^{2}f{f''}+24{f}^{2} \right) g+6r^4f^3 f^{\prime}g^{\prime}-24r^2 f^4=0\,,
\label{const2}
\end{eqnarray}
where, for simplicity, we have defined new constants $\lambda = 8\pi \tilde{\lambda}$ and $\alpha = 8\pi \tilde{\alpha}$.  

We first consider wormhole solutions with zero tidal force, corresponding to a constant redshift function $\Phi(r) = \mathrm{const}$ (whcih implies that $f(r)$ is also constant). 
Thus, substituting $f'(r) = 0$ into the above equation, we obtain the following differential equation:
\begin{eqnarray}
\left[16rg^{\prime}(g-1)-4{r}^{2}(g^{\prime})^{2}-16g^{2}+32g-16\right]\lambda+24r^{2}g+24{r}^{3}g^{\prime}-24{r}^{2}=0\,, 
\label{cons3}
\end{eqnarray}
which yields the solution
\begin{equation}
g(r)=\frac{-4 r^2+8\lambda}{8\lambda}+\frac{ r^2}{8\lambda}\, {\left(2 W\left[\pm \frac{\sqrt{2C_1 \lambda}}{r^3}\right]+2\right)}^2,\label{gWH2}
\end{equation}
where $W[x]$ denotes the principal solution of the Lambert function  that is analytic in $x=0$ \cite{carlo1,vall1}. For real arguments, the Lambert function can be viewed as the inverse of the relation $x = W(x)\,e^{W(x)}$~\cite{matt1}.  Here, we define $W(x)=W_0(x)$ where $W_0(x)$ is referred to as the principal branch of the Lambert function. The
Taylor series of $W_0(x)$ for  $|x| < e^{-1}$ is~\cite{carlo1}
\begin{equation}
W_0(x) = \sum_{n=1}^\infty {(-n)^{n-1} x^n\over n!}.
\end{equation}
To analyze the asymptotic behavior of this solution at spatial infinity, we adopt the approximation
\begin{equation}
g(r)\simeq 1\pm\frac{\sqrt{2C_1}}{\lambda^{1/2} r}+{\mathcal O}\left(\frac{1}{r^4}\right).
\end{equation}
From this, we note that these solutions correspond to an asymptotically flat spacetime. However, the throat condition $g(r_0) = 0$ requires choosing the negative sign in the preceding equation in order to ensure a wormhole solution.  
Imposing $g(r_0) = 0$ then determines the integration constant $C_1$ as
\begin{equation}
C_1=\frac{{r_0}^4\left[-\lambda+r_0\zeta\right]e^{-\frac{2 \zeta}{r_0}}}{\lambda},
\end{equation}
where $\zeta=\left(r_0-\sqrt{r_0^2-2\lambda}\right)$. At the throat, Eq.~(\ref{cons3}) yields
\begin{align}
g^{\prime}(r_0)=\frac{3 r_0^2-2\lambda -3r_0\sqrt{r_0^2-2\lambda}}{\lambda r_0},
\end{align}
so that the flaring-out condition, $g^{\prime}(r_0)>0$, imposes the restriction $r_0^2>2 \lambda$. 

\begin{figure}
	\begin{center}
		\includegraphics[scale=0.412]{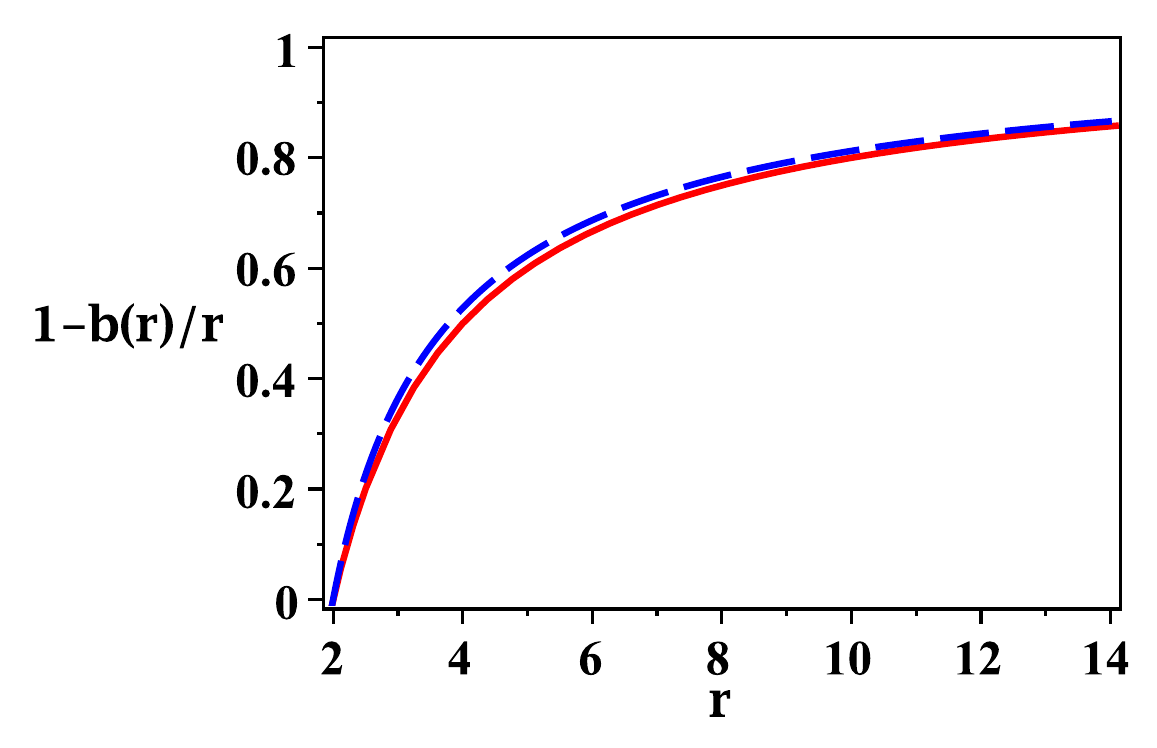}
		\caption{The behavior of  $1-b(r)/r$  with respect to $r$ for $r_0=2$ and $\lambda=0$ (solid red curve), $\lambda=0.5$ (dashed blue curve), respectively.}\label{figbr1}
	\end{center}
\end{figure}

The behavior of $g(r)$ in the small $\lambda$ limit is given by
\begin{equation}
g(r)\simeq 1-\frac{r_0}{r}+\frac{(r^3-r_0^3)}{2r_0 r^4}\lambda+\frac{(r^3-r_0^3)}{2r^7} {\lambda}^2+{\mathcal O}\left({\lambda}^3\right),\label{gaprox}
\end{equation}
which guarantees the asymptotic flatness of the solution. 
Considering different values of the parameter $\lambda$, we plot the function $1 - b(r)/r$ in Fig.~(\ref{figbr1}).  To ensure the asymptotic flatness condition, we obtain $l(r)$ by using the solution (\ref{gaprox}) for small $\lambda$ as
\begin{align}
\label{eq:embedding1}
l(r)=\pm \sqrt{1-\frac{r_0}{r}}\bigg(\frac{6r^2 r_0-2\lambda r_0-5\lambda r}{6r r_0}\bigg) \pm \bigg(\frac{\lambda-2r_0^2}{4 r_0}\bigg) \ln\left[\frac{2r}{r_0}\left(\sqrt{1-\frac{r_0}{r}}+1\right)-1\right],
\end{align}

In order to analyze the energy conditions for this class of solutions, we study the behavior of the quantities $\rho$, $\rho + p_r$, and $\rho + p_t$ throughout the spacetime, which are expressed as
\begin{equation}
\rho=-\frac{3}{2 \lambda}\,
{W\left[-\frac{r_0^2e^{-\frac{\zeta}{r_0}}{[-2\lambda+2 r_0\zeta]}^{1/2}}{ r^3}\right]}^{2},
\end{equation}
\begin{equation}
\rho+p_r=- \frac{1}{\lambda} \, {W\left[-\frac{r_0^2e^{-\frac{\zeta}{r_0}}{[-2\lambda+2 r_0\zeta]}^{1/2}}{ r^3}\right]}\left({W\left[-\frac{r_0^2e^{-\frac{\zeta}{r_0}}{[-2\lambda+2 r_0\zeta]}^{1/2}}{ r^3}\right]}-1\right),
\end{equation}
\begin{equation}
\rho+p_t=-\frac{\left({W\left[-\frac{r_0^2e^{-\frac{\zeta}{r_0}}{[-2\lambda+2 r_0\zeta]}^{1/2}}{ r^3}\right]}\right)\left(3{W\left[-\frac{r_0^2e^{-\frac{\zeta}{r_0}}{[-2\lambda+2 r_0\zeta]}^{1/2}}{ r^3}\right]}+2{W\left[-\frac{r_0^2e^{-\frac{\zeta}{r_0}}{[-2\lambda+2 r_0\zeta]}^{1/2}}{ r^3}\right]}^{2}+1\right)}{2 \lambda \left({W\left[-\frac{r_0^2e^{-\frac{\zeta}{r_0}}{[-2\lambda+2 r_0\zeta]}^{1/2}}{r^3}\right]}+1\right)},
\end{equation}
respectively, and in the asymptotic limit, these quantities reduce to
\begin{equation}
\rho \simeq \frac{3 r_0^4\big[\lambda-r_0 \zeta \big]e^{-\frac{2 \zeta}{r_0}}}{\lambda r^6}+{\mathcal O}\left(\frac{1}{r^9}\right),
\end{equation}
\begin{equation}
\rho+p_r \simeq -\frac{ r_0^2{\big[-2\lambda+2r_0 \zeta \big]}^{1/2}e^{-\frac{\zeta}{r_0}}}{\lambda r^3}+{\mathcal O}\left(\frac{1}{r^6}\right),
\end{equation}
and
\begin{equation}
\rho+p_t\simeq \frac{ r_0^2{\big[-2\lambda+2r_0 \zeta \big]}^{1/2}e^{-\frac{\zeta}{r_0}}}{2\lambda r^3}+{\mathcal O}\left(\frac{1}{r^6}\right).
\end{equation}

\begin{figure}
	\begin{center}
		\includegraphics[width=7cm]{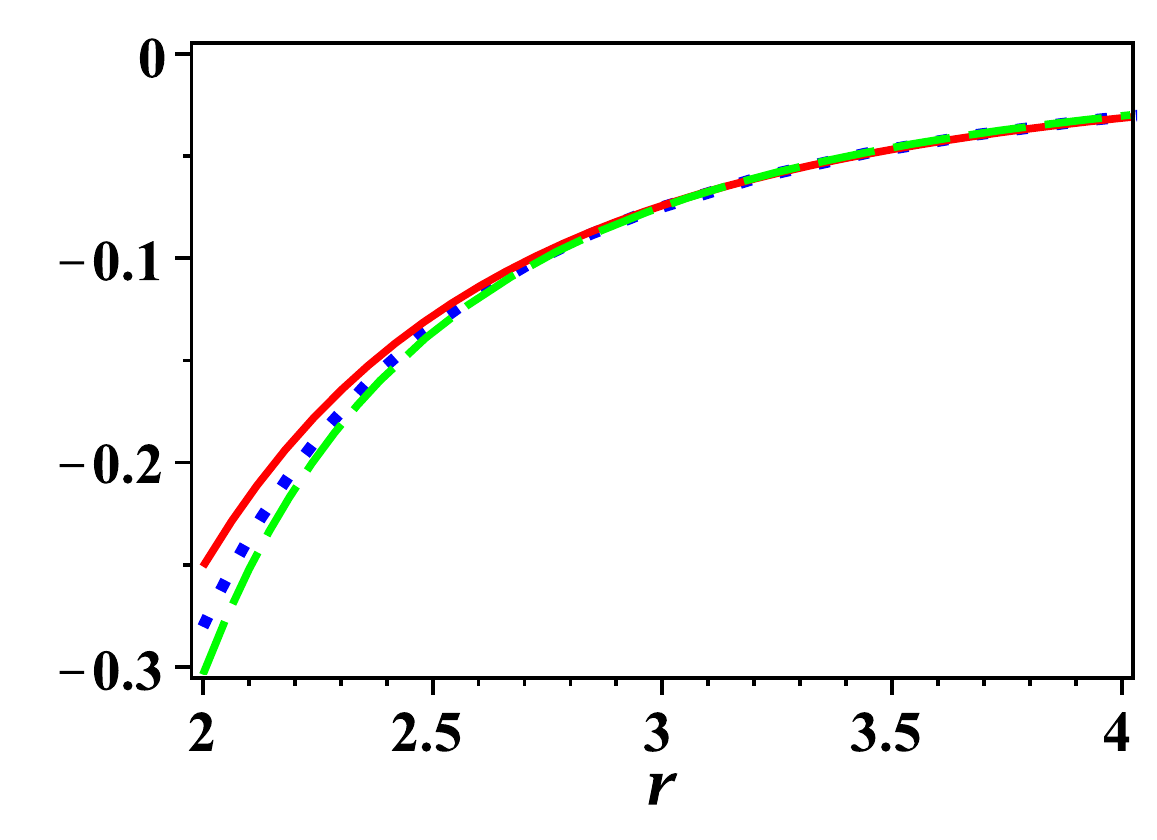}
			\hspace{1cm}
		\includegraphics[width=7cm]{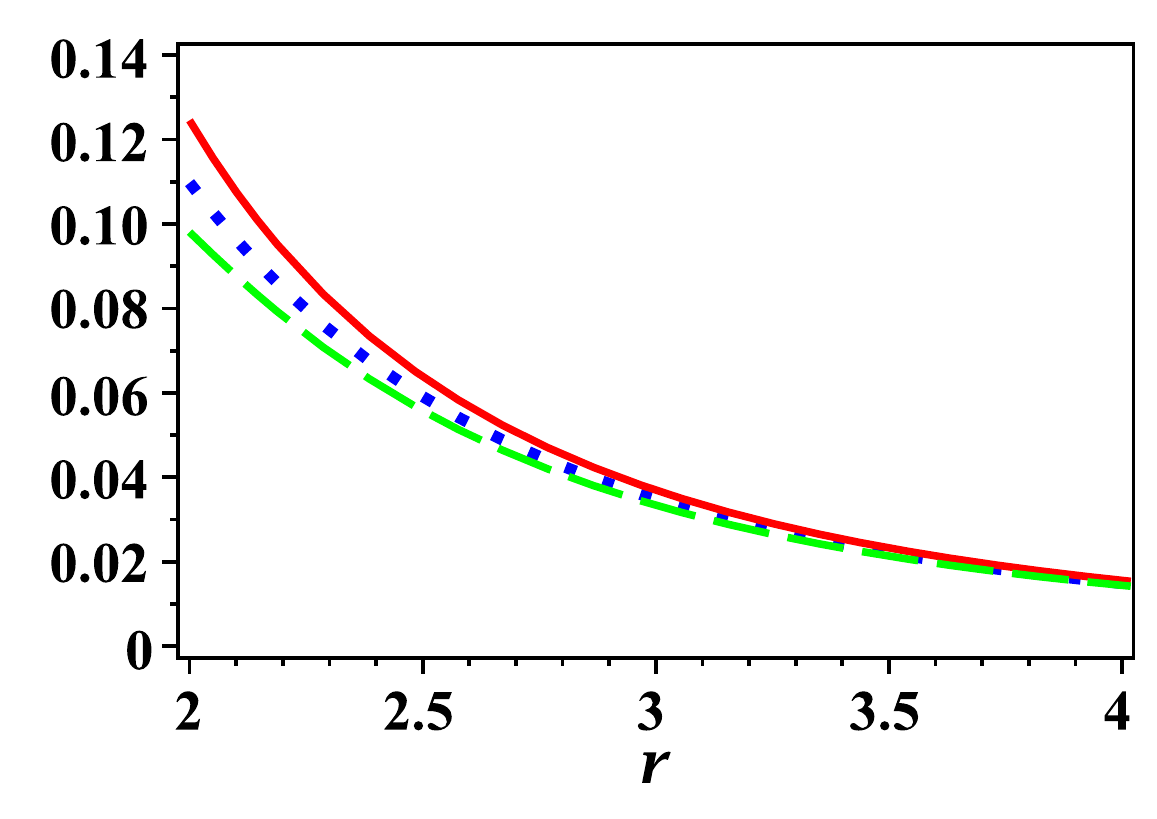}
		\includegraphics[width=7cm]{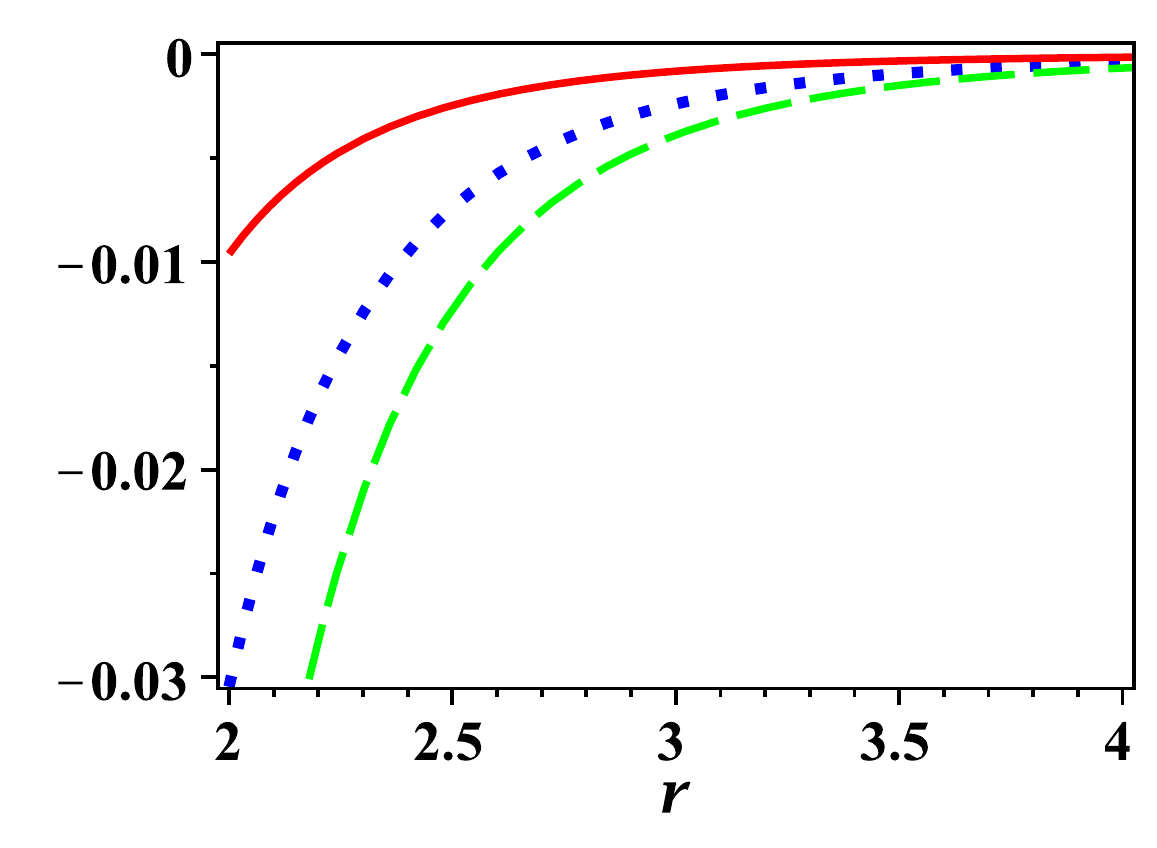}
		\caption{The behavior of  $\protect\rho +p_{r} $ (top left plot), $\protect\rho +p_{t}$ (top right plot) and $\protect\rho $ (bottom plot) versus $r$. The model parameters have been set as, $r_0=2$. The curves from up to down correspond to the cases with $\lambda=0.01$ (red solid curve), $\lambda=0.3$ (blue dotted curve), $\lambda=0.5$ (green dashed curve) respectively.}\label{fwor2}
	\end{center}
\end{figure}

It is evident that both $\rho + p_r$ and $\rho + p_t$ approach zero as $r \to \infty$, but with opposite signs. Consequently, in the large-$r$ limit, one of these quantities becomes negative, leading to a violation of the WEC. The behavior of $\rho$, $\rho + p_r$, and $\rho + p_t$ is illustrated in Fig.~\ref{fwor2}. It is worth noting that all components of the SET tend to zero as $r \to \infty$. Furthermore, $\rho$ and $\rho + p_r$ possess no real roots and remain negative everywhere, with their magnitudes increasing as the parameter $\lambda$ decreases, while $\rho + p_t$ remains positive throughout the spacetime. We therefore conclude that the Ricci scalar and the Kretschmann scalar assume the following finite values at the
throat and correspond to an asymptotically flat spacetime:
\begin{equation}
\mathcal{R}(r)=-\frac{3 r_0^2}{r^6}\lambda+\frac{3r^3-6r_0^3}{r^9} \lambda^2-\frac{3(r^6-12r_0^3 r^3+16r_0^6)}{4 r^{12} r_0^2} \lambda^3+{\mathcal O}(\lambda^4), \label{rics1}
\end{equation}
\begin{equation}
\mathcal{K}(r)= \frac{6 r_0^2}{r^6}-\frac{(r^3-2r_0^3)}{r^9} \lambda+\frac{3(r^6-12 r_0^3 r^3+18r_0^6)}{2 r_0^2 r^{12}} \lambda^2+{\mathcal O}(\lambda^3).
\label{kr1s1}
\end{equation}

\subsection{Specific case: non-constant redshift function}\label{WHS2}

\subsubsection{Curvature coordinates}

Here, we interpret the trace anomaly as corresponding to an effective fluid with a vanishing on-shell energy density. Such a configuration may be relevant in the context of thin-shell wormholes and other exotic compact objects, such as gravastars~\cite{grav1}. Let us further consider an empty-space scenario characterized by $\rho = 0$. Solving the condition $\rho = 0$ in Eq.~(\ref{feild1}) then yields
\begin{align}
g(r)=1-\frac{C_2}{r}. \label{shap2}
\end{align}
In this case, the radius of the throat corresponds to $r=C_2=r_0$. Thus, substituting $g(r)=1-r_0/r$ in the trace anomaly equation~(\ref{const2}) we obtain the following differential equation
\begin{eqnarray}
&4\,f{r}^{2} \left( {r}^{2}\lambda\, \left( r-{r_0} \right) {{f'}}^{2}+ \left( -3{r_0}+2r \right) fr\lambda\,{f'}+3\,{f}^{2} \left( {r}^{3}+\left( 2\lambda-4\alpha \right) {r_0} \right)  \right)  \left( r-{\it r_0} \right) {f''}\notag \\ &-4\,{r}^{4}\lambda\,{f}^{2} \left( r-{r_0} \right) ^{2}{{f''}}^{2}-{r}^{4}\lambda\, \left( r-{r_0} \right) ^{2}{{f'}}^{4}-\left( -12{r_0}+4 r \right) f{r}^{3} \left( r-{r_0} \right) \lambda\,{{f'}}^{3}+4f' f^3 \,\left( 9\lambda-18\alpha \right)r r_0^2\notag \\ &-36\,\lambda\,{{r_0}}^{2}{f}^{4}-{f}^{2} \left( 6{r}^{4}-6{r}^{3}{r_0}+4{r}^{2}\lambda-24\,\alpha\,{r_0}\,r- \left( 3\lambda-24\alpha \right) {{r_0}}^{2} \right) {r}^{2}{{f'}}^{2}+24f^3 f^{\prime} r^5\notag \\ &-18r_0 r^4 f^3 f^{\prime}-24 f^3 r_0 r^2  \left( \lambda-2\alpha \right) f^{\prime}=0.\label{dynamic2}
\end{eqnarray}

Since solving the above differential equation in full generality is rather complicated, we first seek solutions in the presence of the Type~A anomaly only; i.e., we set $\lambda = 0$, so that the contribution from the squared Weyl tensor vanishes. For notational convenience, we define $f(r) = h^2(r)$. Substituting $\lambda = 0$ into the above equation then leads to the following differential equation:
\begin{align}
-6\,h \left( r \right)  \left({r}^{4}-4 r\alpha\,{r_0}\right)  \left( -{r}^{2}+r{r_0} \right) {r}^{2}{h''(r)}+3\,h\left( r \right)  \left( 4{r}^{6}-3\,{r}^{5}{r_0}+8\,{r0}\,\alpha\,{r}^{3}-12\alpha\,{{r_0}}^{2}{r}^{2} \right)h'(r) =0.
\label{dyf3}
\end{align}
The above equation finally leads to the general solution, given by
\begin{equation}
h(r)=h_1\int^{r}_{r_{0}}\frac{r^{\frac{3}{2}}}{\sqrt{(r-r_0)}(-4\alpha r_0+r^{3})}dr+h_0,\label{redshift}
\end{equation}
where $h_0$  and $h_1$ are integration constants. The solution ($\ref{redshift}$) for $\alpha=0$ reduces to 
\begin{align}
h(r)=h_0+h_1\left[\sqrt{(1-\frac{r_0}{r})}\right], \label{intwh}
\end{align}
which is presented in~\cite{viserka}. 

To obtain an analytical solution, we can approximate the integrand of Eq.~(\ref{redshift}) for small values of $\alpha$ parameter as
\begin{align}
h(r)\simeq h_0+h_1\sqrt{1-\frac{r_0}{r}}\bigg[1+4\alpha\left(\frac{\eta}{35r_0^2 r^3}\right)+16\alpha^2 \left(\frac{64 r^3 \eta+231 r_0^6+252 r r_0^5+280 r^2 r_0^4}{3003 r_0^4 r^6}\right)+{\mathcal O}(\alpha^3)\bigg], \label{hh}
\end{align}
where $\eta=16 r^3+8r_0 r^2+6 r_0^2 r+5 r_0^3$. In order to study the behavior of this solution at infinity, we consider the approximation for large $r$, so that we obtain
\begin{align}
h(r)\simeq h_0-h_1\bigg[\frac{1}{r}+\frac{r_0}{4 r^2}+\frac{r_0^2}{8r^3}+{\mathcal O}\left(\frac{1}{r^4}\right)\bigg].
\end{align} 
Note that this solution corresponds to an asymptotically flat spacetime. 

In what follows, we examine the behavior of the quantities $\rho + p_r$ and $\rho + p_t$ both at spatial infinity and at the wormhole throat, in the limit of small $\alpha$. In this regime, we find
\begin{equation}
\rho+p_r \simeq - \frac{\left( {\frac {64}{35}}\,{h_1}\,\alpha+{ h_0}\,{{r_0}}^{2}\right) {r_0}}{\left(\left( { h_0}+{h_1} \right) {{r_0}}^{2}+{\frac {64}	{35}}\,{h_1}\,\alpha \right) {r}^{3}}+{\mathcal O}\left(\frac{1}{r^4}\right),
\end{equation}
and 
\begin{equation}
\rho+p_t \simeq  \frac{\left( {\frac {64}{35}}\,{h_1}\,\alpha+{h_0}\,{{r_0}}^{2}
\right) {r_0} }{2\left(\left( {h_0}+{h_1} \right) {{ r_0}}^{2}+{\frac {64}{35}}\,{h_1}\,\alpha \right) {r}^{3}} +{\mathcal O}\left(\frac{1}{r^4}\right).
\end{equation}
It is clear that both  $\rho+p_r$ and $\rho+p_t$ tend to zero as $r$ tends to infinity, however, with opposite signs. Therefore, in the large $r$
limit, one of $\rho+p_r$ or $\rho+p_t$ is negative and consequently the WEC is violated. Also at throat we have
\begin{equation}
 \rho+p_r\Big|_{r_0}\simeq\left[-\frac{1}{r_0^2}+\frac{{h_1}\,\sqrt {r-{r_0}}}{{h_0}\,{{r_0}}^{5/2}}+{\mathcal O}(r-r_0)+...\right]+\alpha\left[\frac{4h_1\sqrt{\frac{r}{r_0}-1}}{h_0 r_0^4 }-\frac{8\,h_1^{2} \left( r-{r_0} \right)}{r_0^5h_0^2}+...\right]+{\mathcal O}(\alpha^2),\label{wec1}
\end{equation}
\begin{equation}
\rho+p_t\Big|_{r_0}\simeq-\frac{1}{2}\left[-\frac{1}{r_0^2}+\frac{{h_1}\,\sqrt {r-r_0}}{{h_0}\,{{r_0}}^{5/2}}+{\mathcal O}(r-r_0)+...\right]+\alpha\left[\frac{8h_1\sqrt{\frac{r}{r_0}-1}}{h_0 r_0^4 }-\frac{10\,h_1^{2} \left( r-{r_0} \right)}{r_0^5h_0^2}+...\right]+{\mathcal O}(\alpha^2). \label{wec2}
 \end{equation}

\begin{figure}
\begin{center}
\includegraphics[width=7cm]{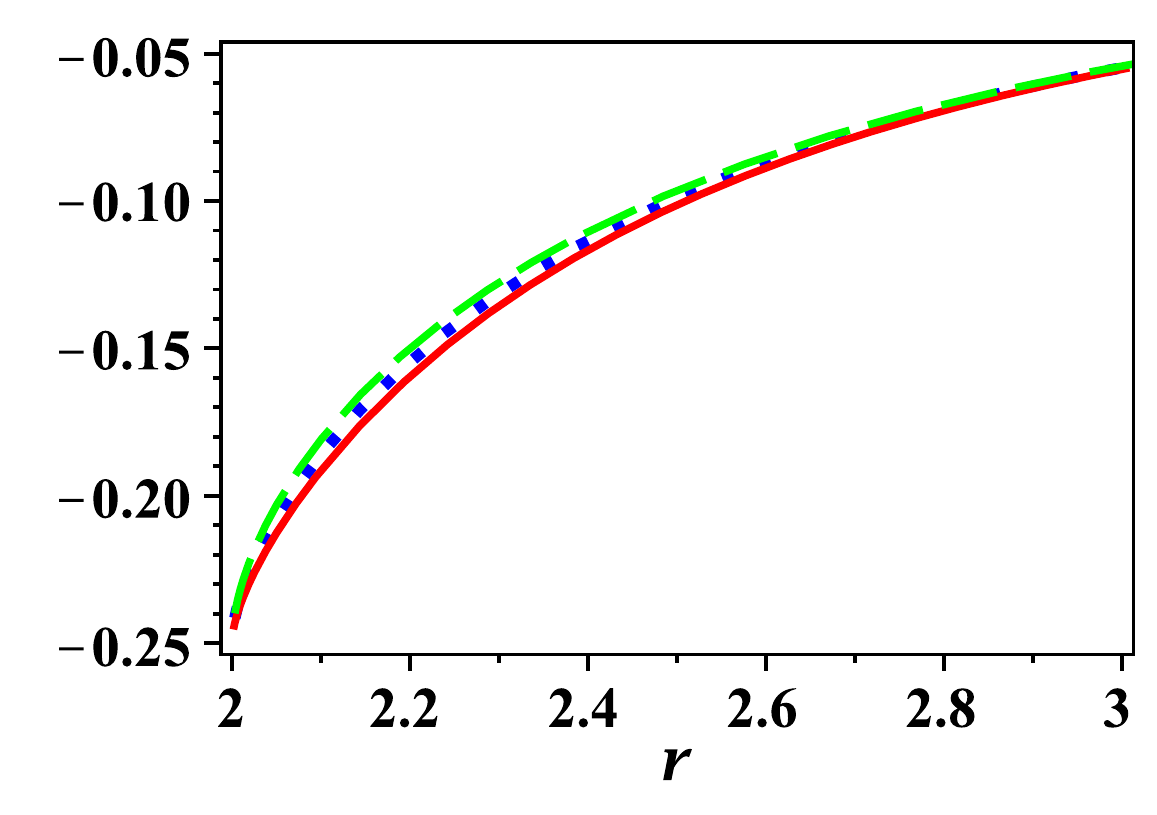}
			\hspace{1cm}
\includegraphics[width=7cm]{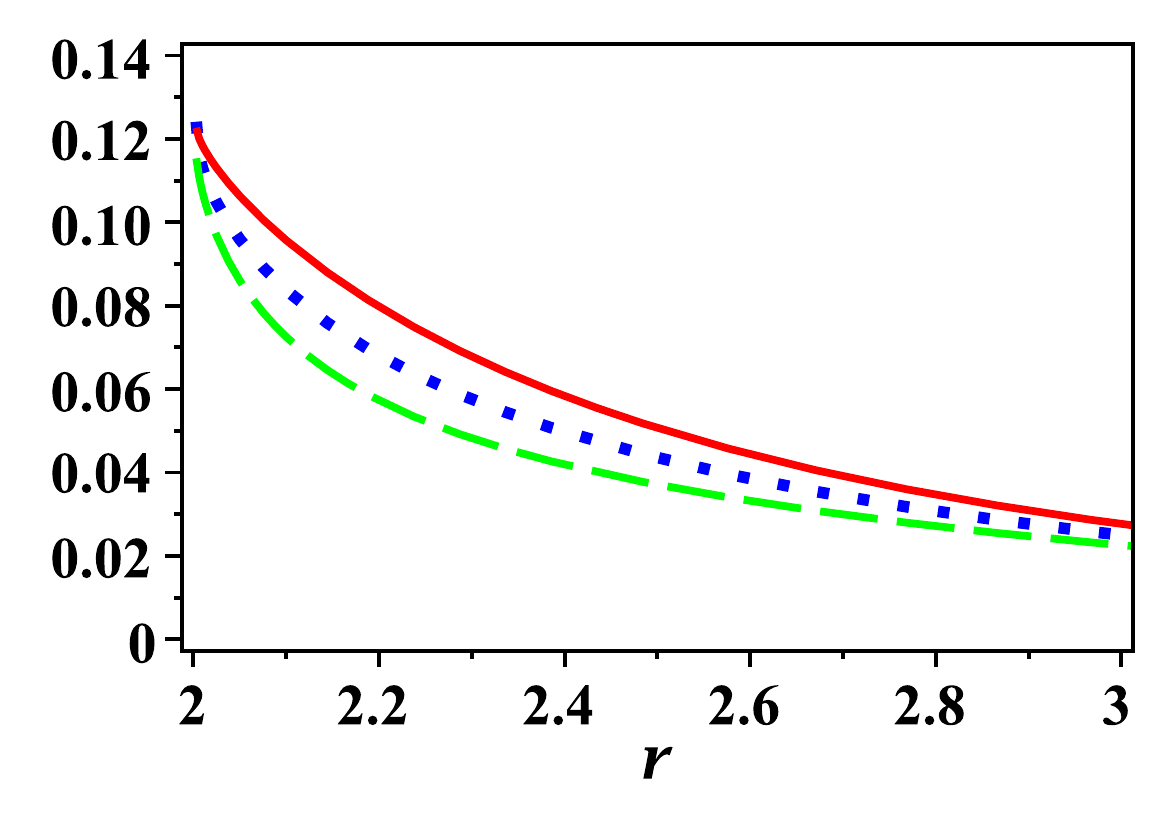}
\caption{The behavior of  $\protect\rho +p_{r} $ (right panel) and $\protect\rho +p_{t}$ (left panel) versus $r$. The model parameters have been set as, $r_0=2$. The solid, dotted and dashed curves correspond to the cases with $\alpha=0,0.03,0.06$ respectively.}\label{fwor3}
\end{center}
\end{figure}

In Eqs.~(\ref{wec1}) and (\ref{wec2}), the first terms, corresponding to the GR case ($\alpha = 0$), have opposite signs, which leads to a violation of the NEC.  
The presence of the trace anomaly introduces a second term with a consistent sign, which can help reduce the amount of exotic matter.  
In Fig.~\ref{fwor3}, we plot the quantities $\rho + p_r$ and $\rho + p_t$ for different values of the parameter $\alpha$.  
We note that the components of the SET vanish as $r \to \infty$.  
Figure~\ref{fwor3} also demonstrates that increasing the value of $\alpha$ reduces the required amount of exotic matter.

\subsubsection{Isotropic coordinates}

To obtain wormhole solutions we must impose several conditions on the metric presented in powers of $\alpha$ with using Eq.~(\ref{hh}). Thus, we have
 \begin{equation}
 ds^{2}=-\left[ h_0+h_1\sqrt{1-\frac{r_0}{r}}\bigg(1+4\alpha\left(\frac{\eta}{35r_0^2 r^3}\right)+{\mathcal O}(\alpha^2)\bigg)\right]^{2}dt^{2}+\left[\frac{dr^{2}}{1-r_0/r} +r^{2}\left(d\theta^{2} + \sin^2\theta \, d \phi^2 \right)\right]\,,\label{metric2}
\end{equation}
It is clear that the general solution~(\ref{metric2}) reduces to the Schwarzschild geometry when $h_0 = 0$ and $\alpha = 0$. Moreover, it contains the spatial-Schwarzschild wormhole as a special case when $h_1 = 0$. In this solution, no event horizon is present; however, a wormhole throat exists at $r = r_0$. A traversable wormhole occurs if, at the throat $r = r_0$ and for all $r > r_0$, no horizon forms. 
To determine appropriate values of $h_0$ and $h_1$ that yield a single global coordinate patch for the traversable wormhole, it is convenient to introduce isotropic coordinates, defined by
\begin{align} 
r={\tilde{r}}{\left(1+\frac{r_0}{4\tilde{r}}\right)}^2.\label{trans1}
\end{align}

We can therefore apply the transformation~(\ref{trans1}) to convert from curvature coordinates to isotropic coordinates, thereby expressing the metric~(\ref{metric2}) in isotropic coordinates to first order in $\alpha$
\begin{equation}
ds^{2}=-\left[h_0+h_1\left(\frac{1-\frac{r_0}{4\tilde{r}}}{1+\frac{r_0}{4\tilde{r}}}\right)-\alpha h_1\left(\frac{64\left(1-\frac{r_0}{4\tilde{r}}\right)\Xi(\tilde{r})}{35 r_0^2 \left(1+\frac{r_0}{4\tilde{r}}\right)^7}\right)\right]^{2}dt^{2}+{\left(1+\frac{r_0}{4 \tilde{r}}\right)}^4\left[d{\tilde{r}}^{2} +{\tilde{r}}^{2}\left(d\theta^{2} + \sin^2\theta \, d \phi^2 \right)\right]\,,\label{metric3}
\end{equation}
where
\begin{align}
\Xi(\tilde{r})=\frac{r_0^6+32\,\tilde{r}r_0^5+464\,{\tilde{r}}^2r_0^4+4096\,{\tilde{r}}^{3}{r_0}^3+7424\,{\tilde{r}}^{4}{{r_0}}^{2}+8192\,{\tilde{r}}^{5}{r_0}+4096\,{\tilde{r}}^{6}}{4096 {\tilde{r}}^6}.
\end{align}

The area of the spherical surfaces defined by $\tilde{r} = \text{constant}$, along with its derivatives, is given by
\begin{align}
A(\tilde{r})=4\pi {\tilde{r}}^2\left(1+\frac{r_0}{4\tilde{r}}\right)^4,
\end{align}
\begin{align}
\frac{dA(\tilde{r})}{d\tilde{r}}=\frac{\pi\left(r_0+4\tilde{r}\right)^3\left(4\tilde{r}-r_0\right)}{32 \tilde{r}^3},
\end{align}
\begin{align}
\frac{d^{2}A(\tilde{r})}{d\tilde{r}^{2}}=\frac{\pi\left(256 \tilde{r}^4+16\tilde{r}r_0^3+3r_0^4\right)}{32 \tilde{r}^4},
\end{align}

One verifies that the area attains its minimum at $\tilde{r} = r_0/4$, and the flaring-out condition is satisfied, namely $A''\!\left(\tilde{r} = \frac{r_0}{4}\right) = 64\pi > 0$. Therefore, the geometry corresponds to a traversable wormhole with the throat located at $\tilde{r} = r_0/4$, in the absence of an event horizon. 
 It is straightforward to verify that the metric (\ref{metric3}) remains invariant under the combined transformation \(\tilde{r} \rightarrow r_0^2/(16\,\tilde{r})\) together with a sign reversal of the parameter \(h_1 \rightarrow -h_1\). This implies that the region \(0<\tilde{r}<r_0/4\) is geometrically equivalent to the region \(\tilde{r}>r_0/4\). Moreover, the limiting value \(\tilde{r}\rightarrow 0\) corresponds to a geometry analogous to that of \(\tilde{r}\rightarrow \infty\). In this sense, the regimes \(\tilde{r}\simeq 0\) and \(\tilde{r}\simeq \infty\) represent the two asymptotically flat ends of the spacetime. Consequently, the geometry on the opposite side of the throat can be generated by applying the transformation \(h_1 \rightarrow -h_1\) to the metric (\ref{metric2}), which takes the isotropic form (\ref{metric3}) via the coordinate transformation (\ref{trans1}).
An event horizon would arise if $g_{tt}$ vanishes, leading to the horizon solution given by
\begin{align}
\tilde{r}_H=\frac{r_0}{4}\frac{h_1-h_0}{h_1+h_0}+\frac{\alpha}{r_0}\frac{-2h_0^3\left(5 h_0^4+35h_1^4-21h_0^2h_1^2\right)+70 h_1^6 h_0}{35{\left(h_0+h_1\right)}^2 h_1^5}. \label{rhoriz}
\end{align}
 To better understand the properties of curvature invariants in these geometries, we can obtain the Ricci scalar~(\ref{ric1}) and the Kretschmann	scalar~(\ref{kr11}) around the region $\tilde{r}_H$ where the $g_{tt}(\tilde{r}_H)=0$ for the global coordinate patch given in Eq.~(\ref{metric3}). We therefore find the following expressions for Ricci and Kretschmann scalars in the limit $\tilde{r} \to \tilde{r}_H$
\begin{equation}
\mathcal{R}(\tilde{r}_H)=\frac{128{\tilde{r}_H}^4{g^{\prime2}_{tt}}(\tilde{r}_H)}{{g_{tt}(\tilde{r}_H)}^2\left(r_0+4\tilde{r}_H\right)^4},\label{rics21}
\end{equation}
\begin{equation}
	\mathcal{K}(\tilde{r}_H) =\frac{16384{\tilde{r}_H}^8 {g^{\prime4}_{tt}}(\tilde{r}_H)}{{g_{tt}(\tilde{r}_H)}^4\left(r_0+4\tilde{r}_H\right)^8},
	\label{kr1s21}
\end{equation}
 whereby we observe that both the curvature scalars diverge as $g_{tt}(\tilde{r}_H) \to 0$ indicating this horizon is actually a naked curvature singularity. However, the curvature singularity does not form, when the $g_{tt}$ component of the metric (\ref{metric3}) never goes to zero, and we therefore have a traversable wormhole. In order to ensure the absence of horizons and singularities, one can choose the parameters characterizing the solution so that $\tilde{r}_H<0$. We first examine the limit $\alpha \to 0$ in Eq. (\ref{rhoriz}), and  find  
\begin{align}
\tilde{r}_H=\frac{r_0}{4}\frac{h_1-h_0}{h_1+h_0},
\end{align}
which implies that there is no horizon or singularity as $\tilde{r}_H<0$. Note that one can choose $h_0$ and $h_1$ as both positive (or negative) with $\lvert{h_0}\rvert>\lvert{h_1}\rvert$ for traversable wormhole solutions. In the present case, due to the presence of the trace anomaly, we define $\xi = h_0/h_1$ in Eq.~(\ref{rhoriz}), so that the condition $\tilde{r}_H<0$ yields the following
\begin{align}
\left(280\xi-280\xi^3+168\xi^5-40\xi^7\right)\alpha-35r_0^2\left(\xi^2-1\right) <0.
\end{align}

We therefore find that the Ricci and Kretschmann scalars take the following finite values at the throat:
\begin{equation}
\mathcal{R}(\tilde{r_0})=\frac{24h_1 \alpha}{r_0^5 h_0}\left(\tilde{r}-\frac{r_0}{4}\right)+{\mathcal O}\left(\tilde{r}-\frac{r_0}{4}\right)^2,\label{rics2}
\end{equation}
\begin{equation}
\mathcal{K}(\tilde{r_0}) = \frac{6}{r_0^4}-\frac{24\left[6 r_0^4 h_0^2-h_1\left(r_0^4+16r_0^2 \alpha+72\alpha^2\right)\right]}{h_0^5 r_0^{10}}\left(\tilde{r}-\frac{r_0}{4}\right)^2+{\mathcal O}\left(\tilde{r}-\frac{r_0}{4}\right)^3.
\label{kr1s2}
\end{equation}

\subsection{General case: $\lambda$ and $\alpha$}
Our objective here is to solve the differential equation~(\ref{dynamic2}) for the shape function introduced in the previous section. We then perform a transformation of the form
	$f(r)={\rm exp}\!\left(2\int W(r)\,dr\right)$ in Eq.~(\ref{dynamic2}), which yields
\begin{eqnarray}
	&4\lambda r^4\left(r-r_0\right)\left[W^{\prime}(r)+W^{2}(r)\right]^2-\left(r-r_0\right)r^2\left(4\lambda\left(2r-3r_0\right)r W(r)+6r^3-12r_0\left(2\alpha-\lambda\right)\right)\left[W^{\prime}(r)+W^{2}(r)\right]\notag \\ &+\lambda r^2\left(2r-3r_0\right)^2 W^{2}(r)-\big(12 r^4-9r_0r^3+12r_0r\left(2\alpha-\lambda\right)-18r_0^2\left(2\alpha-\lambda\right)\big)rW(r)+9\lambda r_0^2=0.\label{eqnew11}
\end{eqnarray}
Since Eq.~(\ref{eqnew11}) cannot be solved analytically for $W(r)$ in terms of standard functions, we first analyze this equation in the vicinity of the wormhole throat. We restrict our attention to the radial region $r_0 < r < r_0 + \Delta$, where $\Delta$ is taken to be infinitesimally small. In this near-throat region, Eq.~(\ref{eqnew11}) can be simplified as follows:
\begin{align}
	r_0^2 \lambda {W(r_0)}^2-3r_0\left(r_0^2+2\lambda-4\alpha\right)W(r_0)+9 \lambda=0, \label{gener2}
\end{align}
which can easily be solved for the value of $W(r_0)$ 
\begin{align}\label{intia1}
	W(r_0)=\frac{3\left(r_0^2+2\lambda-4\alpha\right)\pm3\sqrt{\left(r_0^2+2\lambda-4\alpha\right)^2-4\lambda^2}}{2\lambda r_0}.
\end{align}
As the behavior of the redshift function remains finite in the vicinity of the wormhole throat, we expect the Ricci and Kretschmann scalars to attain finite values at the throat. This behavior can be explicitly verified from Eqs.~(\ref{ric1}) and~(\ref{kr11}) by considering the negative-sign branch of the above solutions
\begin{align}
	\mathcal{R}({r_0})=\frac{6\lambda}{r_0^4}+\frac{24\alpha \lambda -12 \lambda^2}{r_0^6}+{\mathcal O}\left(\frac{1}{r_0^{8}}\right),
	\label{rics3}
\end{align}
and
\begin{eqnarray}
	\mathcal{K}({r_0})\!\!&=&\!\!\frac{6}{r_0^4} +\frac{9\lambda^2}{r_0^8}+\frac{27\alpha \lambda^2}{r_0^{10}}+{\mathcal O}\left(\frac{1}{r_0^{12}}\right).
	\label{kr12s2}
\end{eqnarray}
Also for positive sign solutions the above scalars read
\begin{align}
	\mathcal{R}({r_0})=\frac{6}{\lambda}+\frac{12\lambda-24\alpha}{\lambda r_0^2}+{\mathcal O}\left(\frac{1}{r_0^{4}}\right),
\end{align}
and
\begin{eqnarray}
	\mathcal{K}({r_0})\!\!&=&\!\!\frac{9}{\lambda^2}+\frac{36\lambda-72\alpha}{\lambda^2 r_0^2}+{\mathcal O}\left(\frac{1}{r_0^{4}}\right).
\end{eqnarray}

\begin{figure}
	\begin{center}
		\includegraphics[scale=0.488]{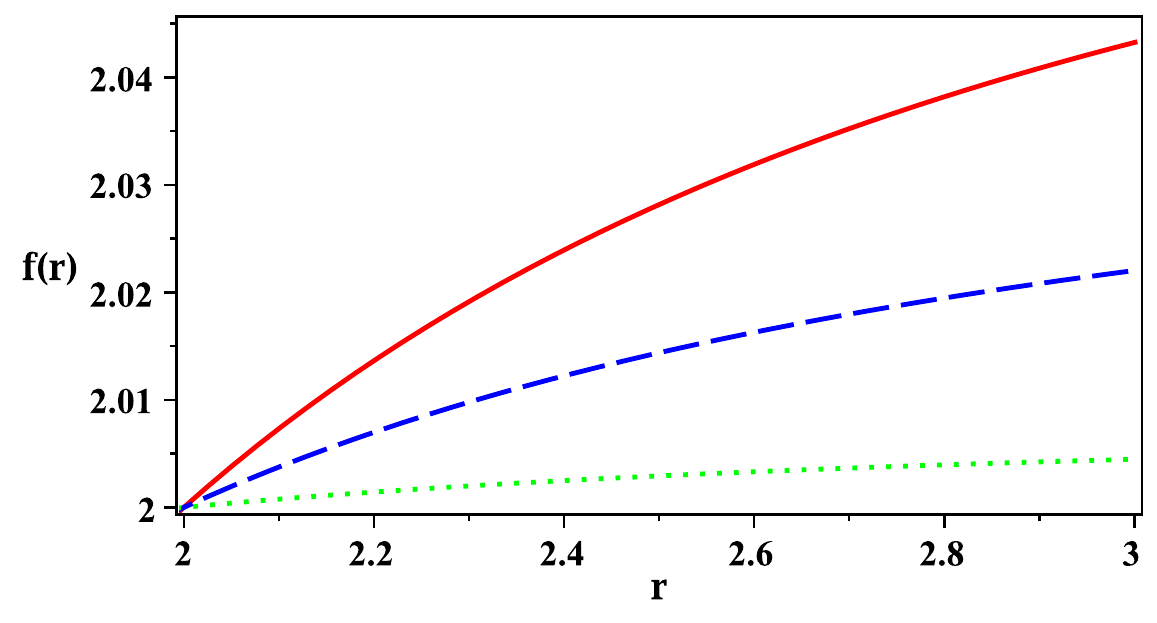}
		\caption{The behavior of $f(r)$ for $\alpha=0.1$ with respect to $r$ for $r_0=2$ and $\lambda=0.01$ (green dashed curve), $\lambda=0.05$ (blue dashed curve), $\lambda= 0.1$ (red solid curve), respectively.}\label{figbr21}
	\end{center}
\end{figure}

It is therefore evident that the Ricci and Kretschmann scalars assume finite values at the throat for the negative-sign branch and remain finite as $\alpha$ and $\lambda$ tend to zero. However, these scalars diverge in the limit $\lambda \to 0$; hence, one may only consider the solutions with the negative sign in Eq.~(\ref{intia1}). As noted above, obtaining exact analytical solutions of Eq.~(\ref{eqnew11}) is extremely challenging. We therefore solve this equation numerically for selected values of the parameters $\alpha$ and $\lambda$ and investigate the corresponding properties of the solutions. In Fig.~\ref{figbr21}, we display the redshift function obtained using the numerical solution of Eq.~(\ref{eqnew11}) for the function $W(r)$, which satisfies the initial condition $W(r_0)$ at the throat. It is worth noting that $W(r_0)$ in Eq.~(\ref{intia1}) with the negative sign yields solutions corresponding to an asymptotically flat spacetime. We observe that $f(r)$ remains finite throughout the spacetime, as expected, indicating the absence of an event horizon in the wormhole geometry. Using Eqs.~(\ref{feild2}) and~(\ref{feild3}), we then obtain the following expressions for the radial and transverse pressures:
\begin{align}
	p_r =\frac{2r\left(r-r_0\right)W(r)-r_0}{r^3},\label{prrnew1}
\end{align}
and
\begin{align}
	p_t =\frac{2r^2\left(r-r_0\right)W^{\prime}(r)+\big(2r\left(r-r_0\right)W(r)+r_0\big)\big(1+rW(r)\big)}{2 r^3}.\label{pttnew1}
\end{align}

\begin{figure}
   	\begin{center}
   		\includegraphics[width=7cm]{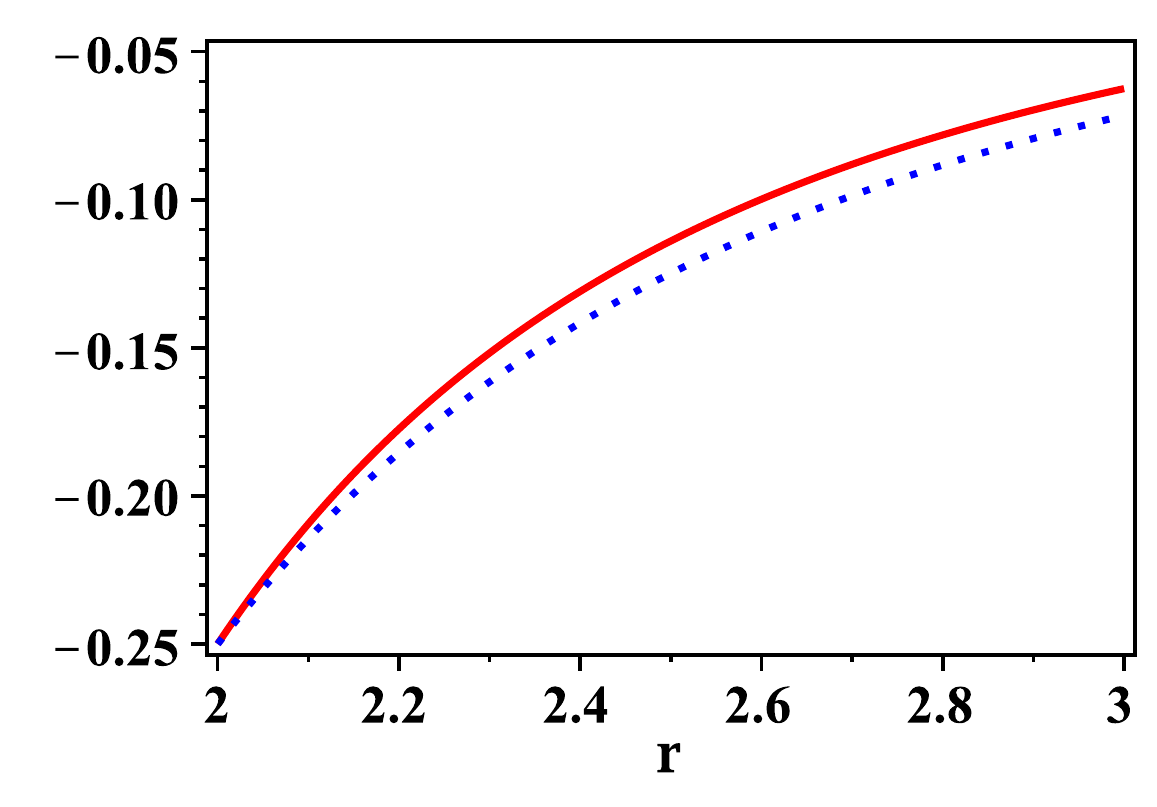}
   					\hspace{1cm}
   		\includegraphics[width=7cm]{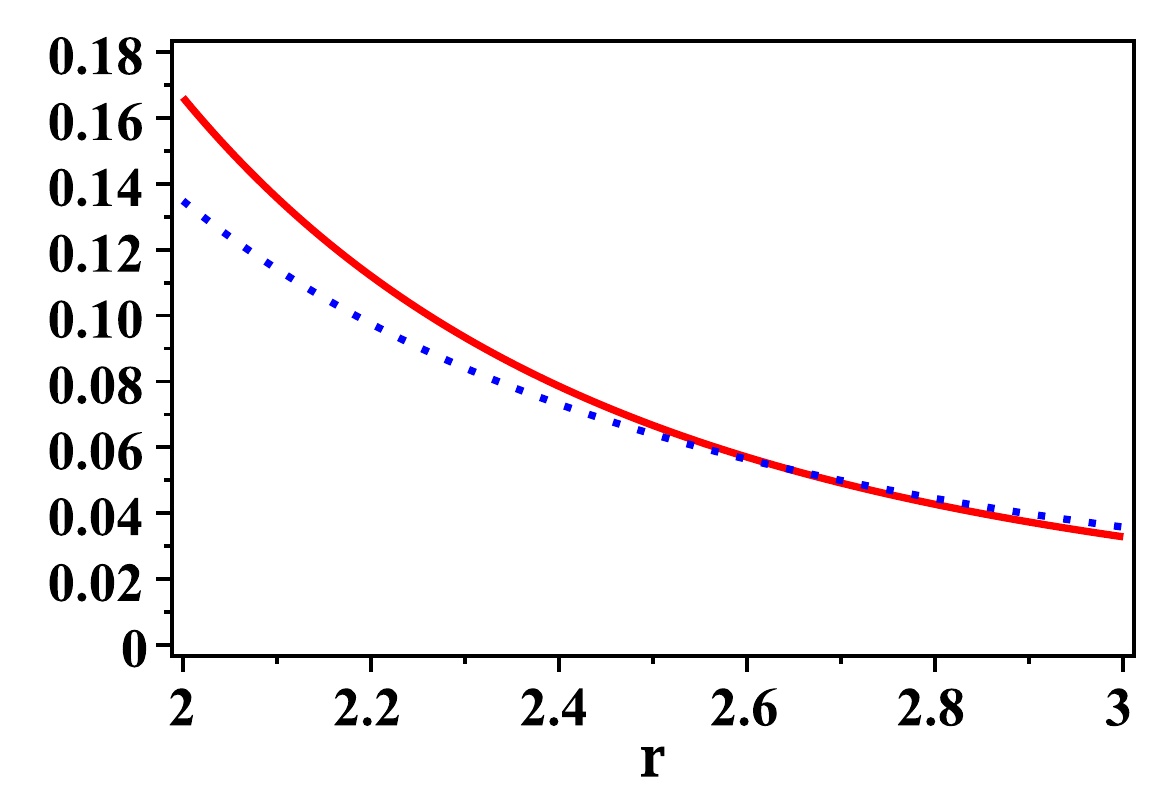}
   		\caption{The behavior of $p_{r}$ (right panel) and $p_{t}$ (left panel) versus $r$. The model parameters have been set as, $r_0=2$ and $\alpha=0.1$. The solid and dotted curves correspond to the cases with $\lambda=0.05, 0.01$ respectively.}\label{wecgenera}
   	\end{center}
   \end{figure}

At wormhole throat, these pressures take the form
\begin{align}
	p_r \simeq\frac{-1}{r_0^2}+\bigg(\frac{3+2r_0W(r_0)}{r_0^3}\bigg)\left(r-r_0\right)+{\mathcal O}\left(r-r_0\right)^2,
\end{align}
and
\begin{align}
	p_t \simeq\frac{1+r_0W(r_0)}{2r_0^2}+\bigg(\frac{2r_0^2W^2(r_0)+3r_0^2{W^{\prime}}(r_0)-3}{2r_0^3}\bigg)\left(r-r_0\right)+{\mathcal O}\left(r-r_0\right)^2.
\end{align}
In Fig.~\ref{wecgenera}, we illustrate the behavior of the radial and tangential pressures as functions of the radial coordinate by employing Eqs.~(\ref{prrnew1})--(\ref{pttnew1}) together with the numerical solutions for the function $W(r)$, for different values of the parameters $\lambda$ and $\alpha$. In particular, the magnitude of the radial pressure is observed to increase in the vicinity of the wormhole throat as the parameter $\lambda$ increases.

\section{Particle Trajectories Around the Wormhole}\label{WHtrajectory}
In this section, we analyze the geodesic equations in the wormhole spacetime described by the metric~(\ref{metric2}) using the Lagrangian formalism~\cite{rinder}. Due to the spherical symmetry, it is sufficient to consider motion confined to the equatorial plane, $\theta = \pi/2$. The corresponding Lagrangian for the metric~(\ref{mor1}) is then given by
  \begin{equation}
  \mathfrak{ L} = g_{\mu\nu} \dot{x}^\mu \dot{x}^\nu= -f(r)\dot{t}^2+\frac{\dot{r}^2}{g(r)}+r^2\dot{\phi}^2   \,,
  \label{lag}
  \end{equation}
where a dot denotes differentiation with respect to the affine parameter $\eta$. 

Since the Lagrangian is constant along a geodesic, we can set $\mathfrak{L}(x^{\mu}, \dot{x}^\mu) = \epsilon$, where time-like and null geodesics correspond to $\epsilon = -1$ and $\epsilon = 0$, respectively. Using the Euler-Lagrange equations
  \begin{equation}\label{lag2}
  \frac{d}{d\eta} \frac{\partial{\frak
  		L}}{\partial\dot{x}^{\mu}}-\frac{\partial{\frak L}}{\partial
  	x^{\mu}}=0,
  \end{equation}
one can readily identify the following constants of motion
  \begin{equation}\label{lag5}
  \dot{t}=\frac{E}{f(r)}\,,  \qquad  r^2\dot{\phi}=L,
  \end{equation}
where $E$ is the energy and $L$ the angular momentum of the test particle.
Inserting these constants of motion into Eq. (\ref{lag}), we get
  \begin{align}\label{radc}
  \dot{r}^{2}=g(r)\left(\frac{E^2}{f(r)}-\frac{L^{2}}{r^{2}}+\epsilon \right).
  \end{align}
 It is convenient to rewrite Eq.~(\ref{radc}) in terms of the proper radial distance $l(r)$ which is finite for all finite values of $r$. Note that the spacetime is extended in such a way that $l$ monotonically increases from $-\infty$ to $+\infty$ so that $l<0$ or $l>0$ correspond to two parallel universes joined together via a throat at $l=0$. Using the proper radial distance, Eq. (\ref{radc}) takes the simple form 
  \begin{align}
  \label{eq:geodesics}
  \dot{l}^{2}=f^{-1}(r)\left[E^{2}-V_{{\rm eff}}(L,l)\right],
  \end{align}
where the effective  potential is defined as
  \begin{align}
  \label{eq:potential}
  V_{{\rm eff}}(L,l)=f\left(r(l)\right)\left(\frac{L^{2}}{r^2(l)}-\epsilon\right).
  \end{align}
  
In what follows, we analyze the trajectories of particles around the wormhole using the above form of the effective potential. In fact, the geodesic equation~(\ref{eq:geodesics}) can be interpreted as a classical scattering problem with a potential barrier $V_{\rm eff}(L, l)$. Furthermore, using the definitions in Eq.~(\ref{lag5}), then Eq.~(\ref{eq:geodesics}) can be rewritten as an ordinary differential equation governing the orbital motion, given by:
  \begin{equation}\label{lag7}
  \left(\frac{dl}{d\phi}\right)^2=\frac{{\dot{l}}^2}{\dot{{\phi}}^2}=\frac{{r^4(l)}}{f(r)L^2}\left[E^2- V_{{\rm eff}}(L,l)\right].
  \end{equation} 

In traversable wormhole spacetimes, particles may cross the throat of the wormhole, moving from one asymptotically flat region of the manifold to another. Accordingly, a geodesic can continue through the throat into the other universe if
  \begin{align}
  E^{2}>V_{{\rm eff}}(L,0).
  \end{align} 
Similarly, for a geodesic reflected back on the same universe by the potential barrier, we have $ E^{2}<V_{{\rm eff}}(L,0)$. In this case, there is a turning point at $l=l_{{\rm tu}}$ which is obtained by solving the following equation 
  \begin{align}
  E^{2}=V_{{\rm eff}}(L,l_{{\rm tu}}).
  \end{align}
  It is easy to verify that
  \begin{align}
  \frac{dV_{{\rm eff}}}{dl}=\sqrt{g(r)}\bigg[\left(\frac{L^2}{r^2}-\epsilon\right) f^{\prime}(r)-\frac{2L^2 f(r)}{r^3}\bigg], \label{derveff}
  \end{align}
and
\begin{align}
  \frac{d^{2}V_{{\rm eff}}}{dl^{2}}=\frac{L^2 f(r)\left(6g(r)-rg^{\prime}(r)\right)}{r^4}+\frac{\big[\left(L^2-\epsilon r^2\right)r^2g^{\prime}(r)-8L^2 r g(r)\big]f^{\prime}(r)}{2r^4}+\frac{\left(L^2-\epsilon r^2\right)g(r)f^{\prime\prime}(r)}{r^2}.
  \end{align}
Substituting the shape function~(\ref{shap2}) into Eq.~(\ref{prrad}), we find
\begin{align}
\label{eq:embedding}
l(r)=\pm r\sqrt{1-\frac{r_0}{r}} \mp \frac{r_0}{2} \ln\left[\frac{2r}{r_0}\left(\sqrt{1-\frac{r_0}{r}}+1\right)-1\right],
\end{align} 
and substituting the redshift function (\ref{hh}) into Eq. (\ref{eq:potential}), we get the following effective potential
\begin{align}
V_{{\rm eff}}(L,l)=\left[ h_0+h_1\sqrt{1-\frac{r_0}{r}}\bigg(1+4\alpha\left(\frac{\eta}{35r_0^2r^3}\right)\bigg)\right]^{2}\left[\frac{L^{2}}{r(l)^{2}}-\epsilon\right]. \label{vef1}
\end{align}
Furthermore, the derivatives of the effective potential~(\ref{vef1}) can be determined using Eq.~(\ref{derveff}) as
\begin{eqnarray}
\frac{dV_{{\rm eff}}}{dl}&=&\sqrt{1-\frac{r_0}{r}}\left[L^2 \alpha^2\left(\frac{4096 r^7-6720r^3 r_0^4-3136r^2 r_0^5-2240r r_0^6}{1225 r^{10}r_0^4}\right)\right]
	\notag \\ 
&&+\sqrt{1-\frac{r_0}{r}}\left[L^2 \alpha\left(\frac{128r^4-20r_0^3 r -96r_0 r^3-32r_0^2 r^2-120r_0^4}{35 r^{7}r_0^2}\right)+L^2\left(\frac{2r(1+{\xi}^2)-3r_0}{r^4}\right)\right]
	\notag \\
	&&+\epsilon\sqrt{1-\frac{r_0}{r}}\left[ \alpha^2\left(\frac{128 r^3+64r_0 r^2+48r r_0^2+40 r_0^3}{35r^8}\right)+\alpha\left(\frac{32r^3+16r^2 r_0+12r r_0^2+80r_0^3}{35 r_0 r^5}\right)+\frac{r_0}{r^2}\right]
		\notag \\
	&&-\xi \left[\frac{2\alpha L^2\left(64r^4-32r_0r^3-8r_0^2 r^2-4r_0^3 r-55r_0^4\right)}{35 r^7}-\epsilon\frac{r_0^3\left(r^3+4\alpha r_0\right)}{2 r^5}+L^2 r_0^2\left(\frac{5r_0-4 r}{2 r^4}\right)\right]. \label{vpr1}
\end{eqnarray}

Next, we study null geodesics ($\epsilon = 0$) for the class of wormhole solutions with a nonzero redshift function~(\ref{redshift}). From Eq.~(\ref{vpr1}), we find two roots satisfying the condition ${V'_{\rm eff}}(r) = 0$ when $\alpha = 0$. These roots are given by $r_1 = r_0$ and $r_2 = \kappa^2 r_0$, where
\begin{align}
{\kappa}^2=\frac{\xi^2-3+\xi\sqrt{\xi^2+3}}{2\left(\xi^2-1\right)}.
\end{align}
Here, we choose $\xi > 1$. Using the second derivative of the effective potential~(\ref{vef1}) at the throat ($r_1 = r_0$), we find ${V''_{\rm eff}}(r) > 0$, indicating a local minimum, whereas at $r_2$ we have ${V''_{\rm eff}}(r) < 0$, corresponding to a local maximum. It is observed that for small values of $\alpha$, we have $\bar{r}_2 = (\kappa^2 + \alpha \bar{r}) r_0$, where
\begin{align}
\bar{r}=-\frac{8\big(\left( 32\,{\kappa}^{6}+6\,{\kappa}^{4}-50\,{\kappa}^{2}+60-112\,{
\kappa}^{8}+64\,{\kappa}^{10} \right) \sqrt {{\kappa}^{2}-1}
+\Sigma_1\big)}{35 \kappa^4 r_0^2\left[\left( 78+ \left( 58\,{\xi}^{2}+58 \right) {\kappa}^{4}+ \left( -56\,{\xi
 	}^{2}-137 \right) {\kappa}^{2} \right) \sqrt {{\kappa}^{2}-1}
 	+\xi \kappa\, \left( 135-251\,{\kappa}^{2}+116\,{\kappa}^{4} \right)\right]},
 \end{align}
and
\begin{align}
\Sigma_1=\kappa\, \left( -51\,{\kappa}^{2}+55+4\,{\kappa}^{4}-96\,{\kappa}^{8}
+24\,{\kappa}^{6}+64\,{\kappa}^{10} \right) \xi.
\end{align}

\begin{figure}
	\begin{center}
		\includegraphics[scale=0.448]{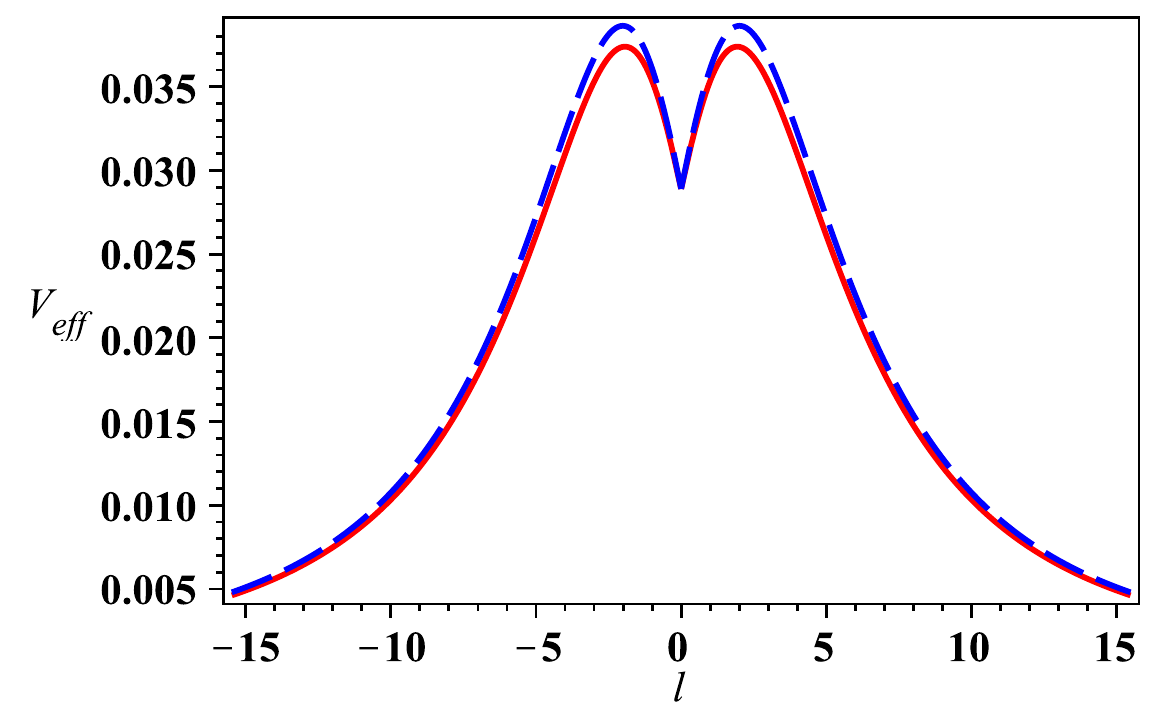}
		\caption{Effective potential for null geodesics for $\alpha=0$ (solid curve) and $\alpha=0.2$ (dashed curve) with $r_0=3$ and $L=0.5$.}\label{veffnuul1}
	\end{center}
\end{figure}

It can be readily shown that the second derivative of the effective potential at $\bar{r}_2$ is positive. In Fig.~\ref{veffnuul1}, we illustrate the behavior of the effective potential as a function of the proper radial distance for different values of the parameter $\alpha$. We observe that both the height and depth of the effective potential increase as $\alpha$ increases.  At the throat a discontinuity in the derivative of the effective potential is observed which can be due to gluing two symmetric spacetimes through a thin shell, see e.g.~\cite{thinshell} for more details. It is worth mentioning that for  $\alpha=0$, the metric in Eq. ~(\ref{metric2}) reduces to the wormhole solutions ~\cite{viserka}, whose gravitational lensing was investigated in Ref.~\cite{Shaikh2019}.

For massive particles, the innermost stable circular orbits (ISCO) for the metric~(\ref{metric2}) can be determined. To find the circular orbits (denoted by $r_c$), one substitutes $\epsilon = -1$ into Eq.~(\ref{vpr1}) and solves $V'_{\rm eff}(L, r_c) = 0$. However, there is no simple analytic expression for $r_c$ in terms of $r_0$ and $\xi$. Therefore, we consider Eq.~(\ref{vpr1}) for $V'_{\rm eff}(L, r_c) = 0$, which leads to the critical angular momentum $L_c(r_c, r_0, \xi)$, given by
\begin{align}
L_c(r_c,r_0,\xi)=
{\bigg(-\frac{7r_0^3 r_c^2\left[r_c^3+4\alpha r_0 \right]\big(175\,r_c^{3} \left( \xi\,\sqrt {r_c}+\sqrt{r_c-r_0}\right){{r_0}}^{2}+\alpha\sqrt{r_c-r_0}\Sigma_2\big)}{{\alpha}^2\sqrt{r_c-r_0}\big(128\,r_c \left( -64\,r_c^{6}+105\,r_c^{2}r_0^{4}+49\,r_cr_0^{5}+35\,r_0^{6} \right) +3600 r_0^7\big)+\Sigma_3}\bigg)}^{\frac{1}{2}}, \label{angul}
\end{align}
where
\begin{align}
\Sigma_2=\big[40r_c(8r_c^2+4r_cr_0+3r_0^2)+100r_0^3\big],
\end{align} 	
and
\begin{eqnarray}
\Sigma_3&=&140\alpha r_0^2 r_c^3\left[\xi\left(\sqrt {r_c} \left( 55\,r_0^{4}+4\,r_0^{3}r_c+8\,r_c^{2}r_0^{2}+32\,{r_0}\,{r_c}^{3}-64\,{r_c}^{4} \right)\right)\right]
	\notag \\ 
	&& + 280\alpha r_0^2\,r_c^4 \left( 5\,r_0^{3}+8\,r_c{{r_0}}^{2}+24\,{r_c}^{2}{r_0}-
  	32\,{r_c}^{3} \right) \sqrt{r_c-r_0}
  		\notag \\
  	&& + \left(-2450\, \left( 1+{\xi}^{2} \right) r_0^{4}{r_c}^{7}+3675\,r_0^{5}r_c^{6}\right)\sqrt{r_c-r_0}+1225 \xi r_0^4 \sqrt{r_c}\left(5 r_0r_c^6-4r_c^7\right).
  	\end{eqnarray}
The energy of this spacetime responsible for circular orbits can be calculated using the above expressions. 

\begin{figure}
 	\begin{center}
 		\includegraphics[width=7cm]{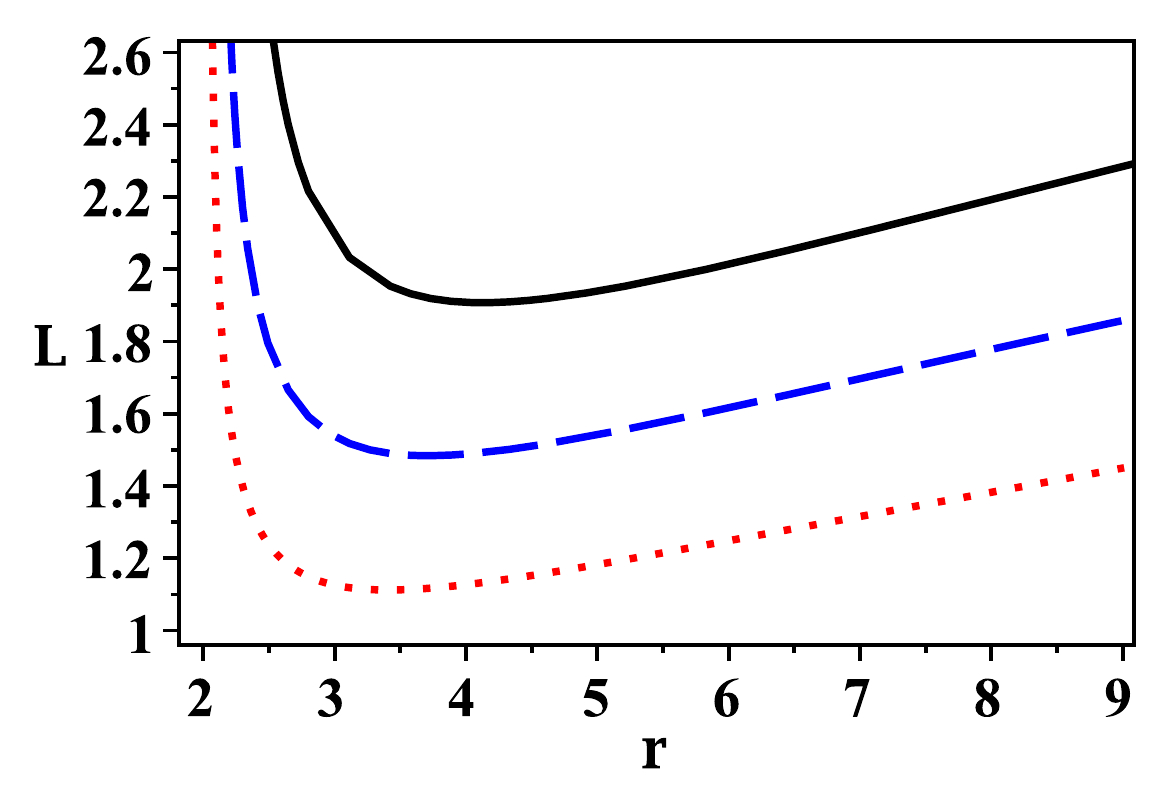}
 		   					\hspace{1cm}
 		\includegraphics[width=7cm]{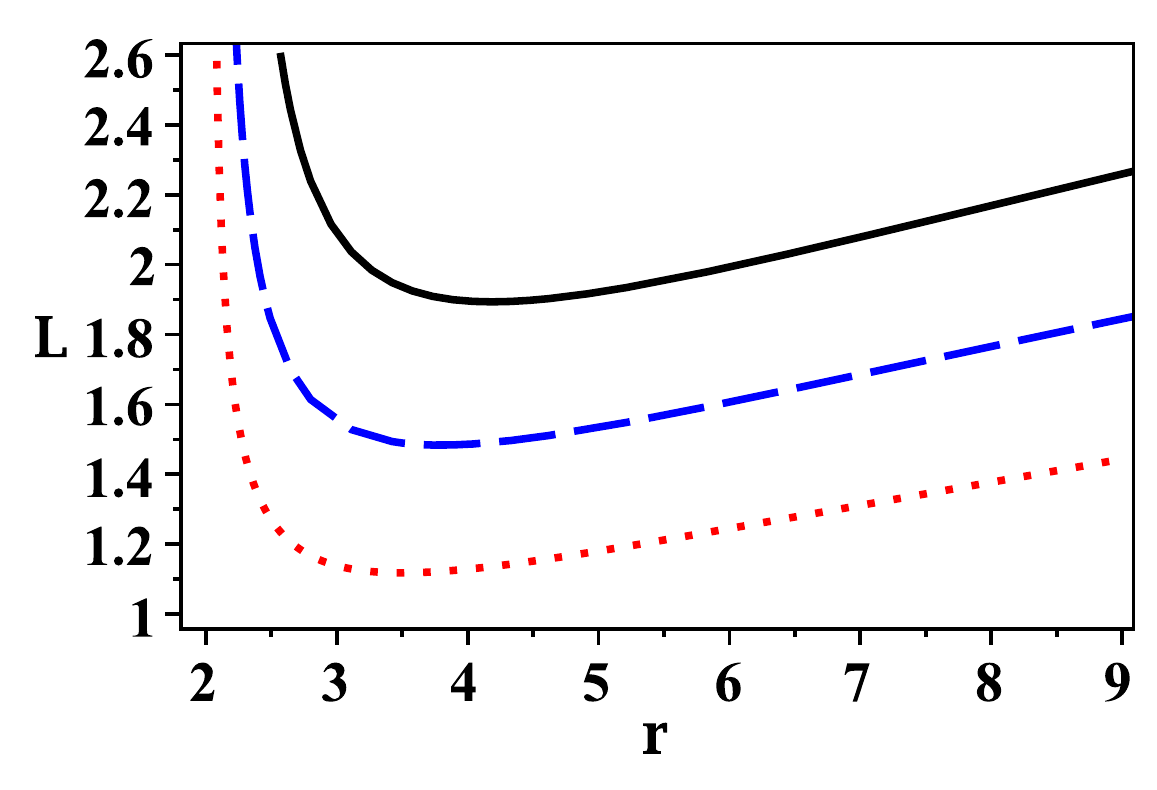}
 		\caption{Radial dependence of the angular momentum for particle on circular orbits. The left pplot is related to the case of $\alpha=0$ and the right one corresponds to the case of $\alpha=0.1$. The solid, dashed and dotted curves correspond to the cases with $\xi=1.2,2.2,4.1$ respectively with $r_0=2$.}\label{Lversur}
 	\end{center}
 \end{figure}

In Fig.~\ref{Lversur}, we show the radial dependence of the angular momentum for a particle on circular orbits. It is evident that the circular orbits shift toward larger radii $r$ as the parameter $\alpha$ increases. The ISCO radius of a test particle can be determined by imposing the auxiliary condition $V''_{\rm eff}(L, r_c) = 0$ and using Eq.~(\ref{angul}) to obtain a first-order approximation in the parameter $\alpha$
\begin{eqnarray}
&&\left( -210\,{\xi}^{3}+70\,\xi \right)r_0^{2}r_c^{11/2}+\left(-70-210\,{\xi}^{2} \right) r_0^{2}\sqrt {{r_c}-{r_0}}r_c^{5}+\left( 280+420\,{\xi}^{2} \right)r_0^{3}\sqrt {{r_c}-{r_0}}r_c^{4}
  \notag \\ &+& \left(  \left( 140{\xi}^{3}-280\xi \right)r_0^{2} \left({r_c}-{r_0} \right) + \left( 245{\xi}^{3}-35\,\xi \right) r_0^{3} \right) r_c^{9/2}+\left( -35r_0^{4}\xi+490r_0^{3}\xi \left( {r_c}-{r_0} \right)  \right) r_c^{7/2} \notag\\ &-& 210r_c^{3}r_0^{4}\sqrt {{r_c}-{r_0}}+\alpha \left[256\xi r_c^{11/2}+ \left( -384{\xi}^{2}-384 \right) r_c^{5}\sqrt{{r_c}-{r_0}}+ \left( -128\,{r_0}\,\xi-1024\,\xi\, \left( {r_c}-{r_0}
\right)  \right)r_c^{9/2}\right] \notag\\
&&+\alpha\left[\left( 280+420{\xi}^{2} \right) {{r_0}}^{3}\sqrt {{r_c}-{r_0}}{{r_c}}^{4}+ \left( -35r_0^{4}\xi+490r_0^3\xi\left( {r_c}-{r_0} \right)\right)r_c^{7/2}-210r_c^{3}r_0^{4}\sqrt {{r_c}-{r_0}}
\right]=0. 
\label{linear1}
  \end{eqnarray}
  
Since solving the above equation analytically is, in general, highly complicated, we impose restrictions on the parameter $\alpha$ and consider the case of small values of this parameter. First, setting $\alpha = 0$ in Eq.~(\ref{linear1}) and simplifying, we obtain
\begin{align}
(\xi^2-1) r_c^3+(7-3\xi^2)r_0 r_c^2+\left(\frac{9}{4}\xi^2-15\right)r_0 r_c +9r_0^3=0\label{cubic1}
\end{align} 
By solving the cubic equation~(\ref{cubic1}), the general solution is given by 
\begin{align}
r_c=\frac{(3\xi^2-7)r_0}{3(\xi^2-1)}+\delta+u {\delta}^{-1},
\end{align}
where
\begin{align}
\delta={\big(v+\sqrt{v^2-u^3}\big)}^{1/3},
\end{align}
\begin{align}
v=-\frac{(495\xi^2+54 \xi^4+27 \xi^6-64) r_0^3}{216 (\xi^2-1)^3},
\end{align}
\begin{align}
u=\frac{(9 \xi^6+30 \xi^4-23 \xi^2-16)r_0^2}{36 (\xi^2-1)^3}.
\end{align}

It is readily seen that in the limit $\xi = 0$, the circular orbit radius is $r_c = 3 r_0$~\cite{alvis}. Since obtaining general analytic solutions is extremely difficult, we adopt a numerical approach to solve Eq.~(\ref{linear1}). In doing so, we select the constant $\xi$ such that wormhole solutions are obtained. To this end, the value of $\xi$ is determined to first order in the parameter $\alpha$ as
 \begin{align}
 \xi=\frac{2\sqrt{r_c(r_c-r0)}(r_c-3r_0)}{r_c(3r_0-2r_c)}-\frac{8\alpha \sqrt{r_c(r_c-r_0)}\Sigma_4}{35(3r_0-2r_c)^2 r_c^4 r_0^2}, \label{xx1}
 \end{align}
 where
 \begin{align}
 \Sigma_4=32r_c^{5}-80r_c^{4}{r_0}+36r_c^{3}r_0^{2}+10r_c^{2}r_0^{3}-423\,{r_c}r_0^{4
 }+495r_0^{5}.
 \end{align}
 
\begin{figure}
  	\begin{center}
  		\includegraphics[scale=0.498]{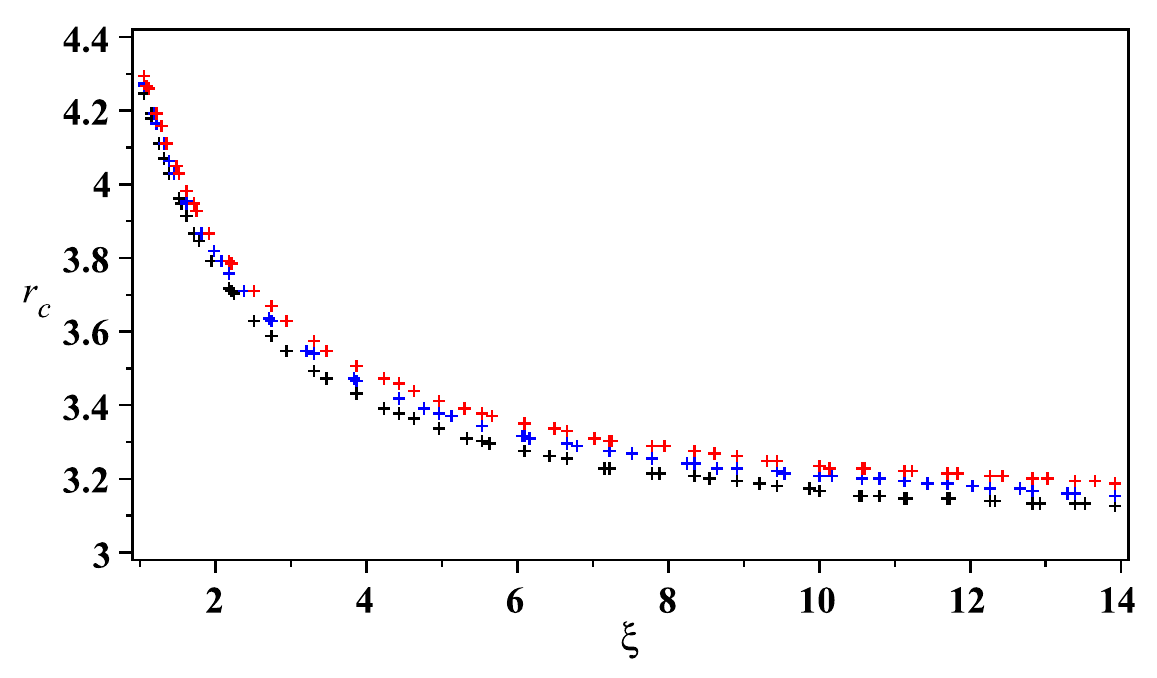}
  		\caption{The behavior the ISCO radius $r_c$ versus $\xi$. The black, blue and red  curves correspond to the cases with $\alpha=0, 0.05 , 0.1$ respectively with $r_0=2$.}\label{rcversuxi1}
  	\end{center}
  \end{figure}

In Fig.~\ref{rcversuxi1}, we plot the ISCO radius obtained by numerically solving Eq.~(\ref{xx1}) for different values of the parameter $\alpha = 0, 0.05, 0.1$ as a function of $\xi$, with $r_0 = 2$. As shown in Fig.~\ref{rcversuxi1}, the ISCO radius increases monotonically due to the presence of trace anomaly corrections.


\section{Summary and concluding remarks}\label{concluding}

In this work, we have investigated a novel class of wormhole solutions arising within the framework of the semi-classical Einstein’s equations in the presence of trace-anomaly effects. The corresponding geometries are supported by a SET sourced by the conformal anomaly, which effectively provides the exotic matter content necessary to sustain a traversable throat. These solutions are characterized by two positive parameters, $\alpha$ and $\lambda$, which encode both the matter content of the underlying quantum theory and the number of degrees of freedom associated with the quantum fields. By systematically analyzing the limiting cases of Type A ($\lambda=0$) and Type B ($\alpha=0$) anomalies, we have been able to construct a broad spectrum of spacetime geometries, ranging from Lorentzian wormholes to naked singularities and Schwarzschild black holes. Moreover, our study identifies the regions in parameter space where physically consistent and traversable wormhole solutions can exist, thereby illustrating the crucial role of trace-anomaly contributions in shaping the global structure of semi-classical spacetimes.

For the case of a Type B anomaly ($\alpha = 0$), we constructed wormhole solutions under the assumption of a constant redshift function and demonstrated that the components of the SET exhibit a monotonic growth with the parameter $\lambda$, reflecting the increasing contribution of anomaly effects to the spacetime geometry. In the complementary case of a Type A anomaly ($\lambda = 0$), we generalized previously known solutions~\cite{viserka}, thereby obtaining a broader family of spacetimes that encompasses Lorentzian wormholes, naked singularities, and the Schwarzschild black hole as particular limits. Furthermore, by employing isotropic coordinates, we were able to identify explicit parameter ranges that lead to physically viable traversable wormhole geometries, characterized by a regular and well-defined throat structure.

We further extended our analysis to incorporate the full contribution of the trace anomaly, solving the corresponding differential equation in the vicinity of the throat in order to explicitly determine the redshift function. These generalized solutions reveal that the Ricci and Kretschmann scalars remain finite at the throat, thereby confirming the regularity of the geometry and the absence of curvature singularities at the wormhole core. In addition, we observed that both the radial and transverse pressures increase with growing values of $\lambda$, highlighting the sensitivity of the wormhole’s stress-energy distribution to the anomaly parameters and emphasizing the role of quantum effects in shaping the local matter content required to sustain the throat.
 
Finally, we carried out a detailed analysis of particle dynamics in these wormhole geometries, focusing on both null and timelike geodesics. For massless particles, we found that the height and width of the effective potential increase monotonically with the anomaly parameter $\alpha$, indicating that stronger anomaly contributions enhance the gravitational lensing effects and modify the stability properties of photon orbits near the throat. For massive particles, our study showed that the innermost stable circular orbit (ISCO) radius also grows with increasing $\alpha$, reflecting the fact that trace-anomaly effects push stable orbits outward and thereby influence accretion processes and orbital dynamics in the vicinity of the wormhole. These results demonstrate that anomaly parameters play a central role not only in shaping the wormhole geometry itself but also in governing the motion of test particles within these exotic spacetimes.

These results highlight the profound impact that trace-anomaly effects can have on the structure of wormhole spacetimes, showing that quantum corrections are capable of significantly modifying both the underlying geometry and the motion of particles in their vicinity. In particular, the anomaly parameters not only determine the stress-energy content required to support the wormhole throat, but also control key dynamical features such as the effective potential for null trajectories and the innermost stable circular orbit for massive particles. Overall, our study demonstrates that trace-anomaly corrections provide a versatile and robust framework for constructing traversable wormholes with finite curvature invariants and adjustable physical properties. This establishes a deeper connection between quantum field theoretic effects and semi-classical gravity, offering new insights into the nature of exotic spacetimes and potentially guiding future explorations of quantum-gravitational phenomena in astrophysical and cosmological contexts.

\section*{Acknowledgements}
FSNL acknowledges support from the Funda\c{c}\~{a}o para a Ci\^{e}ncia e a Tecnologia (FCT) Scientific Employment Stimulus contract with reference CEECINST/00032/2018, and funding through the research grants UIDB/04434/2020, UIDP/04434/2020 and PTDC/FIS-AST/0054/2021.



\end{document}